\documentclass[twocolumn,tighten]{aastex631}
\usepackage{amsmath}

\newcommand{\eb}{\begin{equation}}
\newcommand{\ee}{\end{equation}}

\usepackage{subfigure,graphicx}
\usepackage{url}
\usepackage{color}
\definecolor{darkgray}{gray}{0.4}
\definecolor{patinared}{rgb}{.72,.10,0}
\definecolor{patinablue}{rgb}{0,.20,.65} 
\definecolor{orange}{rgb}{1,0.5,0}
\definecolor{rkka}{RGB}{219,66,32}

\hypersetup{
  colorlinks=True,      
  urlcolor=darkgray,    
  linkcolor=patinared,  
  citecolor=patinablue, 
}

\shorttitle{VLBI Position Variability I}
\shortauthors{Cigan et al.}
\graphicspath{{./}{figures/}}

\received{June 12, 2024}
\revised{July 19, 2024}
\accepted{July 22, 2024}

\submitjournal{ApJS}

\begin{document}

\title{Metrics of Astrometric Variability in the International Celestial Reference Frame: \protect\\
I. Statistical analysis and selection of the most variable sources}

\email{phillip.j.cigan.civ@us.navy.mil}

\author[0000-0002-8736-2463]{Phil Cigan}
\affiliation{U.S. Naval Observatory, 3450 Massachusetts Ave NW, Washington, DC 20392-5420, USA}

\author[0000-0003-2336-7887]{Valeri V.\ Makarov}
\affiliation{U.S. Naval Observatory, 3450 Massachusetts Ave NW, Washington, DC 20392-5420, USA}

\author[0000-0002-4902-8077]{Nathan J.\ Secrest}
\affiliation{U.S. Naval Observatory, 3450 Massachusetts Ave NW, Washington, DC 20392-5420, USA}

\author[0000-0001-8009-995X]{David Gordon}
\affiliation{U.S. Naval Observatory, 3450 Massachusetts Ave NW, Washington, DC 20392-5420, USA}

\author[0000-0002-4146-1618]{Megan C.\ Johnson}
\affiliation{U.S. Naval Observatory, 3450 Massachusetts Ave NW, Washington, DC 20392-5420, USA}

\author[0000-0001-6759-5502]{Sebastien Lambert}
\affiliation{SYRTE, Observatoire de Paris, Universit\'{e} PSL, CNRS, Sorbonne Universit\'{e}, LNE, 61 avenue de l’Observatoire 75014 Paris,
France}



\begin{abstract}

Using very long baseline interferometry data for the sources that comprise the third International Celestial Reference Frame (ICRF3), we examine the quality of the formal source position uncertainties of ICRF3 by determining the excess astrometric variability (unexplained variance) for each source as a function of time.  We also quantify multiple qualitatively distinct aspects of astrometric variability seen in the data, using a variety of metrics.  
Average position offsets, statistical dispersion measures, and coherent trends over time as explored by smoothing the data are combined to characterize the most and least positionally stable ICRF3 sources. 
We find a notable dependence of the excess variance and statistical variability measures on declination, as is expected for unmodeled ionospheric delay errors and the northern hemisphere dominated network geometries of most astrometric and geodetic observing campaigns. 

\end{abstract}

\keywords{Astrometry(80), VLBI(1769),}


\section{Introduction} \label{sec:intro}

At the turn of the millennium, thirty years of developments in our understanding of extragalactic radio sources and the use of the very long baseline interferometery (VLBI) technique culminated in the 1997 adoption by the International Astronomical Union (IAU) of the International Celestial Reference Frame \citep[ICRF;][]{1998AJ....116..516M}, the realization at radio wavelengths of the International Celestial Reference System \citep[ICRS; e.g.,][]{1995A&A...303..604A}. The ICRF, comprised of the VLBI positions of distant active galactic nuclei (AGNs), supplanted the preceding sequence of star-based catalogs including the Fifth Fundamental Catalogue \citep[FK5;][]{1988VeARI..32....1F,1991VeARI..33....1F}, Hipparcos \citep{1997A&A...323L..49P}, and FK6 \citep{2000VeARI..37....1W}. The ICRF has since undergone two major updates. ICRF2, replacing ICRF1 in 2010, saw a five-fold increase in the number of reference frame objects and selection of ``defining'' sources based primarily on position stability, achieving a dramatic improvement in source position uncertainties and axis stability \citep{2009ITN....35....1M,2015AJ....150...58F}. ICRF3 \citep{2020A&A...644A.159C}, adopted as the fundamental realization of the ICRS in 2019,\footnote{\url{https://www.iau.org/static/resolutions/IAU2018_ResolB2_English.pdf}} saw several major improvements. First, higher priority was placed on having defining sources uniformly distributed across the sky, resulting in a largely different set of defining sources from ICRF2. Second, the slow change in the galactocentric acceleration vector of the solar system, which induces a $\sim5$\,$\mu$as~yr$^{-1}$ aberration drift in extragalactic source positions \citep[e.g.,][]{2003A&A...404..743K,2006AJ....131.1471K,2011A&A...529A..91T}, was modeled and accounted for. Finally, while ICRF1 and ICRF2 were defined in the standard geodetic $S$/$X$ bands (2.3/8.4~GHz, respectively), ICRF3 includes $K$ and $X$/$Ka$ versions (24~GHz and 8.4/32~GHz, respectively), making it the first multi-wavelength realization of the ICRS.

At visual (optical) wavelengths, the reference frame realized by the European Space Agency (ESA) Gaia astrometric mission \citep[Gaia-CRF3;][]{2022arXiv220412574G} became the official instantiation of the ICRS at the beginning of 2022,\footnote{\url{https://www.iau.org/static/archives/announcements/pdf/ann21040c.pdf}} replacing the venerable Hipparcos reference frame \citep{1997ESASP1200.....E} and for the first time producing an optical reference frame of precision comparable to the ICRF. The unprecedented precision of Gaia astrometry has been a mixed blessing, however, as previous hints that the apparent optical and radio positions of many ICRF objects do not quite agree \citep{2002AJ....124..612D,2013MNRAS.430.2797A,2013A&A...553A..13O,2014AJ....147...95Z} were brought into sharp relief \citep{2017A&A...598L...1K,2017ApJ...835L..30M,2017MNRAS.467L..71P,2018AJ....155..229F,2019ApJ...873..132M,2019MNRAS.482.3023P,2020A&A...634A..28L}. Given that the ICRF objects are by selection ``radio-loud'', the natural interpretation of these offsets is that they have something to do with differences in the apparent optical and radio positions of AGN jets. Indeed, for VLBI sources with detected linear, jet-like radio features, there is a clear correlation between the position angle of the linear feature and that of the optical-radio offset \citep{2017A&A...598L...1K,2019MNRAS.482.3023P,2019ApJ...871..143P}, and optical-radio offsets have been found to be correlated with radio source structure more generally \citep{2021A&A...647A.189X}.  On the optical side, astrometry of AGNs is perturbed at the milliarcsecond level by extended structures of nearby host galaxies, gravitational microlensing effects, and optical and physical multiplicity \citep{2012MmSAI..83..952M,2022ApJ...933...28M}.

While the radio and optical instantiations of the ICRS have seen dramatic improvements over the last few decades, it is clear that we are now in an era where the parsec-scale astrophysics of AGNs, especially as it pertains to their multi-wavelength and time-dependent morphologies, plays a regular and critical role in reference frame work. This has been the central focus of the Fundamental~Reference~AGN~Experiment \citep[FRAMEx;][]{2020jsrs.conf..165D}, which has so far focused on spatially resolved studies of the inner parsec of nearby AGNs using the Very~Long~Baseline~Array (VLBA) with supporting data from other facilities \citep{2021ApJ...906...88F,2022ApJ...927...18F,2022ApJ...936...76S}. At moderate redshifts, relativistic AGN jets can appear to exhibit motions significantly in excess of $c$ for a range of jet angles with respect to the line-of-sight (LOS), due to special relativity. With the $\sim8$\,pc\,mas$^{-1}$ angular scale of a typical $z=1$ source,\footnote{Where needed, we assume a flat $\Lambda$CDM cosmology with $h=0.7$ and $\Omega_\mathrm{M}=0.3$.} motion on the scale of a few tens of $\mu$as is expected on the timescales of VLBI monitoring campaigns. Moreover, changes in the synchrotron opacity of the compact radio core due to, e.g., a flare may introduce ``core shift'' variability \citep[e.g.,][]{2019MNRAS.485.1822P}, and the apparent position of the core itself is frequency-dependent, varying at the $0.1-1$~mas level for VLBI sources \citep[e.g.,][]{2011A&A...532A..38S, 2012A&A...545A.113P}. 

If the radio positions of reference frame objects are dominated by a jet (as opposed to a compact, flat-spectrum core), and the optical positions are skewed significantly away from the compact AGN accretion disk by this jet, then this presents a major challenge for the ICRF and instantiations of the ICRS more broadly. Indeed, methods for optimal alignment of reference frames at different wavelengths (e.g., by setting source weights) and accounting for wavelength and time-dependent source positions are a central focus of a recently established IAU working group.\footnote{\url{https://www.iau.org/static/science/scientific_bodies/working_groups/329/charter_icrf-multiwaveband-wg.pdf}} Weighting the ICRF according to source physics may indeed be possible in the immediate term. This was recently suggested by \citet{2022ApJ...939L..32S}, who showed that photometric variability information from Gaia can be used to dramatically reduce the rate of optical-radio offsets, and more broadly argued that photometric variability can be used to predict the likelihood of a source developing an optical-radio offset in the future, creating a means of optimally weighting ICRF sources.

Despite the growing body of work examining photometric and positional variability of ICRF and VLBI objects, to date there has not been a systematic examination of VLBI position variability of the ICRF as a whole. This is a major deficiency, as the catalog positions of ICRF objects are generally a mean value over one or more decades of VLBI sessions, with the sessions for some objects spanning nearly 40 years \citep[in the $S$/$X$ bands;][]{2020A&A...644A.159C}. The time-averaged positions of the ICRF objects may therefore be in significant disagreement with their single-epoch positions (such as those used for spacecraft navigation), or their positions averaged over much shorter timescales (such as those from Gaia). An analysis quantifying the time-dependent positions of ICRF objects is therefore of utmost importance to reference frame work, and is likely also informative for studies of AGN jet physics and variability. Indeed, in a recent paper \citet{LambertSecrest2024} show that, for a sample of 520 ICRF3 sources, VLBI position dispersion is inversely correlated with optical photometric variability, supporting the conclusion of \citet{2022ApJ...939L..32S} that bona fide, line-of-sight blazars should be preferred for a positionally stable ICRF.

In this work, we present the first systematic analysis of VLBI source position variability of the ICRF. In Section~\ref{sec:data}, we give an overview of the VLBI sessions used, and how their data were handled.  In Section~\ref{sec:distributionstats}, we discuss formal statistics and calculations relating to the distributions of VLBI position offsets in some detail, and discuss some of the systematics that may affect our findings.  In Section~\ref{sec:analysis}, we discuss the methods we use to analyze the various categories of position variability. Our results are presented in Section~\ref{sec:results}, wherein we quantify the level of excess (unexplained) position variance in each source, correlations between excess variance and observational systematics (e.g., source declination), corrections for systematic-induced excess variance, and the nature and functional form of intrinsic (astrophysical) position variance. We also compare position variability with other measures of variability, such as radio and optical fluxes, and we quantitatively examine the role of VLBI position variability in optical-radio position offsets. Our main conclusions are listed in Section~\ref{sec:conclusions}, after which we provide some qualitative recommendations for accounting for radio position variability in the context of the ICRF. Finally, a catalog of excess position covariances and corrections is supplied for ICRF3.

\section{Data} \label{sec:data}

VLBI astrometric source positions are regularly determined through repeated observations using global networks of antennae over many years, from which the analyzed group delays are then combined and positions solved for in a global solution \citep[for a recent review, see][]{2022Univ....8..374D}.
Diurnal sessions are scheduled regularly by groups including the IVS, using networks of stations to create baselines of hundreds to thousands of kilometers long, to observe many radio sources multiple times over the course of 24 hours.  
Group delays for the sources and baselines are determined for each observation in a session by scientists at various analysis centers worldwide, including the U.S. Naval Observatory.
The observations considered here use the standard simultaneous X/S band setup, where the S-band at 2.3~GHz is used for determining the ionosphere contribution to the delay, and improving the final solution defined at 8.4~GHz in the X-band.

We generated astrometric source position time series from 6581 of these diurnal sessions spanning back to 1980 using the software package Calc/Solve, which is maintained by NASA Goddard Space Flight Center (GSFC).\footnote{\url{https://space-geodesy.nasa.gov/techniques/tools/calc_solve/calc_solve.html}}
Parameters such as source and station positions, are solved for either ``globally'' (a single resulting value) or as ``arc'' parameters (estimates derived for each measurement).
The solutions also solve for daily Earth orientation parameters and sub-daily troposphere delays and clock polynomials.
The solution is tied to the ICRF3 frame by imposing a no-net-rotation constraint on the 
303 ICRF3 defining sources from their official ICRF3 catalog values.  

A series of five global solutions was executed, each in a similar manner to the standard procedure, but with a subset of one fifth of the source positions being solved for on a session-by-session basis as arc parameters, and the remaining sources solved for globally.  
In this manner, iterating through five subsets of sources, a full time series of source positions in R.A.\ and declination, including their associated errors, was constructed for all ICRF3 source measurements. 
Quarterly global solution products and time series data are available for download at the USNO web pages.\footnote{\url{https://crf.usno.navy.mil/quarterly-vlbi-solution}}

It should be noted that the dataset used here has a few differences from the exact dataset used to generate ICRF3 in 2018 at GSFC. First, there are some early databases that have become unusable after conversion to a new database format. Second, many of the databases were analyzed independently at USNO and may differ slightly in terms of editing and parameterization. And third, approximately five additional years of X/S data have been collected since ICRF3 and are incorporated here.

Current USNO quarterly X/S global solutions combine measurements from over 5550 radio sources, but in the present work we only consider the 4536 sources constituting the ICRF3 catalog.  
To ensure meaningful statistics, we further restrict our considered sample to only include sources that were observed in at least 4 diurnal sessions, with at least 10 total observed delays, and with observations that span a minimum of 2 years. 
Thus, the following work is conducted on 4334 sources drawn from the full ICFR3 catalog, including 299 of the defining sources.

\section{Distributions of VLBI astrometric measurements}
\label{sec:distributionstats}

The collection of astrometric position measurements analyzed in this paper includes $322\,013$ single-epoch determinations for 5162 radio sources. The number of observations per object is quite uneven, with the modal value 5 and the greatest number 4876. Only 938 sources have more than 10 sessions. Eight sources have only one position determination---these are discarded from further analysis along with all individual measurement that have rank-deficient covariance matrices (i.e., have rank 1) or have condition numbers above 100. The latter criterion is meant to avoid numerically unstable results for nearly degenerate data.\footnote{Data points with poorly conditioned or rank-deficient covariances can still be utilized in the ML mean position calculation, with a little more toil.}  The mathematical condition number quantifies sensitivity to perturbations in the data, with extremely high values rendering a matrix ill-conditioned.  For a covariance matrix, the condition number equals the square of the ratio of the major and minor axes of the corresponding confidence ellipse. Thus, this filter removes all extremely elongated or degenerate (when $\rho_i=1$) error ellipses. The remaining number of sources is 5153 with $312\,769$ single-epoch determinations.

\subsection{Testing the distribution of normalized position offset magnitudes}
\label{test.sec}

For compactness in the following discussion of position offsets and statistics, we now use the general and intuitive variables $x$ and $y$ to refer to the single-epoch right ascension and declination components, respectively.
The normalized single-epoch position offset
\eb
D=\sqrt{(x-\bar x,y-\bar y)\,
\boldsymbol C^{-1}\,(x-\bar x,y-\bar y)^T}
\label{D.eq}
\ee
is a scalar variate, which is expected to be Rayleigh-distributed with scale 1 for normally distributed residuals $x-\bar x$, $y-\bar y$, and accurate covariances $\boldsymbol C^{-1}$. It provides the simplest means to test if the observed sample distribution of $\{x,y\}$ positions is consistent with the
assumption of binormal distribution, Eq. \ref{pdf.eq}. There is a complication, however, arising from the fact that the estimated mean position $\{\bar x,\bar y\}$ for each source is the sample mean, not the true position. This effectively removes two degrees of freedom from the $D$ statistics for each source, or one observation. The resulting sample distribution of $D$ values underestimates the true dispersion. To minimize this bias, we limit our analysis here to $288\,853$ position measurements of 921 sources with more than 10 measurements.

\begin{figure*}
\includegraphics[width=0.48 \textwidth]{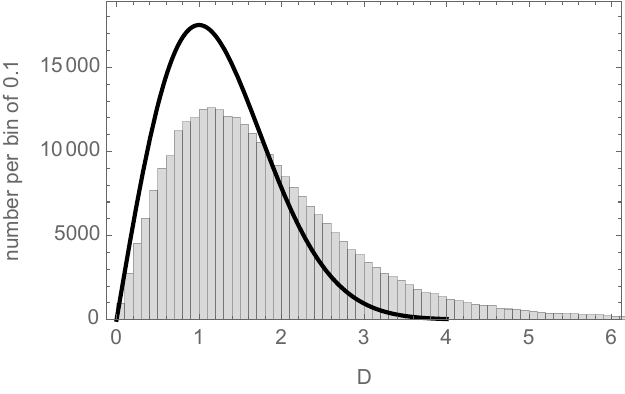}
\includegraphics[width=0.48 \textwidth]{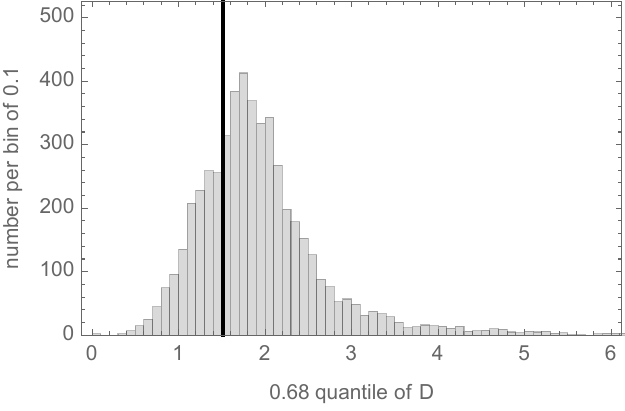}
\caption{Left: Histogram of normalized position offset magnitudes $D$ (Eq. \ref{D.eq}) for VLBI sources with more than 10 measurements. The expected Rayleigh distribution with scale 1 is shown as a black curve. \newline
Right: Histogram of 68th percentiles of normalized offsets $D$ for 5370 VLBI sources. The black vertical line indicates the expected 68th percentile for a Rayleigh[1] distribution.}
\label{D.fig}
\end{figure*}

Fig. \ref{D.fig} shows the histogram of thus computed normalized offset magnitudes and, for comparison, the expected Rayleigh$[1]$ distribution. Using the Anderson-Darling hypothesis test, we estimated the probability ($p$-value) that the two distributions come from the same statistical population and obtained zero. There are two obvious differences between the distributions. The top of the histogram is depleted because of the powerful tail stretching into the domain of high $D$ values, which should be improbable for Gaussian-distributed PDFs. This is a fairly common situation with astronomical measurements, however. The deficit of values around the mode is caused by the heavy tail of the sample distribution stretching much beyond range of finite probabilities. The peak is slightly shifted from 1.

Based on the parameter $D$ computed for each single-epoch observation, we introduce a robust statistical measure of astrometric ``noisiness'' for each source $Q68$, which is defined as the 0.68-quantile (or 68th percentile) of the sample distribution of $D$. The expected value of $Q68$ within the null hypothesis, which is the 0.68 quantile of Rayleigh[1], equals 1.51. Therefore, sources with $Q68$ values in excess of 1.51 are more astrometrically noisy or perturbed than should be expected from the given formal uncertainties. Effectively, this parameter is a robust alternative to the standard deviation of $D$ values for a given source. Fig. \ref{D.fig}, right, shows the histogram of $Q68$ computed for the entire working sample of 5153 VLBI sources. The black vertical line marks the expected value 1.51. The sample distribution is peaked at a higher value than 1.51. A thin positive tail indicates that there are sources with most of the collected measurements strongly perturbed with respect to the formal uncertainty. While we find 27\% of the sources to have $Q68$ below 1.51, 6\% are above 3, and 0.5\% are above 10. These estimates of the noisiness may be underestimated for sources with small $n_{\rm obs}$ by the loss of degrees of freedom.

\subsection{Testing the distribution of standardized position offset vectors}

\begin{figure}
\includegraphics[width=\linewidth]{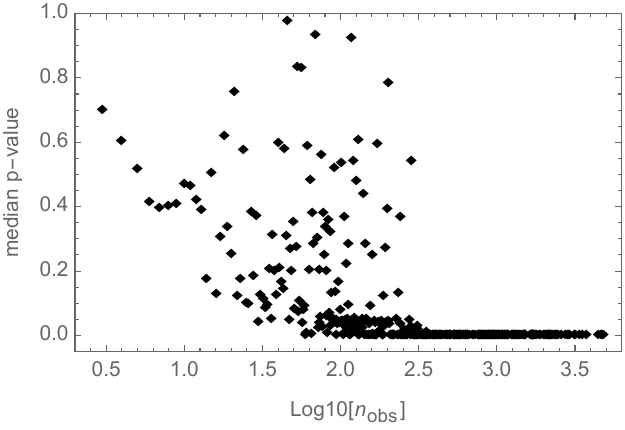}
\caption{Median $p$-values of the standard binormal distribution test versus decimal logarithm of $n_{\rm obs}$ for 5138 VLBI sources with more than 2 position measurements.}
\label{medp.fig}
\end{figure}

Using the standardized position offset vectors with respect to the computed maximum likelihood (ML) positions (Section \ref{sta.sec}, Eq. \ref{st.eq}), more specific analysis of the underlying distribution can be performed, including the remaining covariance not captured by the formal errors. The null hypothesis to be tested is that the vectors $\boldsymbol\delta_i$ for a given source come from the standard binormal distribution (Eq. \ref{pdfsta.eq}). We implemented hypothesis tests for 5138 sources with more than 2 available measurements, using a total of $312\,741$ position determinations and the combined Mardia's method \citep{Mardia1970}. The output $p$-values were further sorted in discrete bins by the corresponding $n_{\rm obs}$, and the median $p$-values computed for each bin. These median $p$-values versus $\log \,n_{\rm obs}$ are depicted in Fig. \ref{medp.fig}. We find that the null hypothesis is never accepted for $n_{\rm obs}>300$. Sources with a small number of data points, $n_{\rm obs}>10$, are often compliant with the binormal hypothesis and the given formal covariance. The median $p$-value of sources with $10<n_{\rm obs}<300$ is equal to 0.05, with only 10\% of their number having $p$-values above 0.52. We conclude that the frequently observed ICRF sources almost never follow the basic statistical model assumed in the data processing algorithms.

We can further test if the coordinate measurements $\{x,y\}$ become uncorrelated upon standardization of variables. We computed the Spearman correlation (commonly known as Spearman $\rho$) between the components of vectors $\boldsymbol\delta_i$ for each of 387 sources with $n_{\rm obs}>100$. The full range of Spearman correlations is between $-0.43$ and $+0.44$, with 275 sources (71\%) having values within $\pm 0.1$. Thus, we find a significant fraction of the source sample with nonzero correlations between the standardized coordinates. We visually inspected the measured positions of several ICRF3 sources with the largest absolute values of Spearman correlations. In all these cases, the distribution of single-epoch positions is visibly elongated in directions not captured by the formal covariances. The sources with the greatest Spearman $\rho$ values in excess of 0.4 are IERS B 0430+052, 1038+52A, 1451$-$375, and 2353$-$686.

\pagebreak

\subsection{Fitting distributions of standardized position offsets}
\label{fit.sec}
We have seen evidence that the sample distributions of measured positions with respect to the ML mean positions are often inconsistent with the implicit assumption of binormally distributed errors specified by the formal covariances. Our goal is now to determine the character of actually measured position distributions. To this end, the powerful distribution fitting methods can be employed, which are available for univariate samples. The distribution-fitting routines also require sufficiently large samples to produce reliable results. Limiting this analysis to 86 ICRF3 sources with more than 1000 single-epoch measurements, best-fitting distributions were uniformly computed with free distribution parameters.\footnote{Wolfram Mathematica function \texttt{FindDistribution} was used to fit sample distributions, \url{https://reference.wolfram.com/language/ref/FindDistribution.html}} The emerging distributions of $D$ values on these large samples are mostly Gamma or Fr\'{e}chet distributions, sometimes with a mixture of Lognormal, Extreme Value, or Cauchy distributions. The location parameters of the fitted Fr\'{e}chet distributions are always negative. A negative location parameter is a technical feature caused by concave shape of the sample histograms in the lowest bins, i.e., a locally positive second derivative near zero. This can be interpreted as a zone of avoidance around the mean position, where the number density of data points is lower than a normal bell-shaped distribution would have. The shape parameters of the Gamma distributions are always above 2 but below 3.

\begin{figure}
\includegraphics[width=0.48 \textwidth]{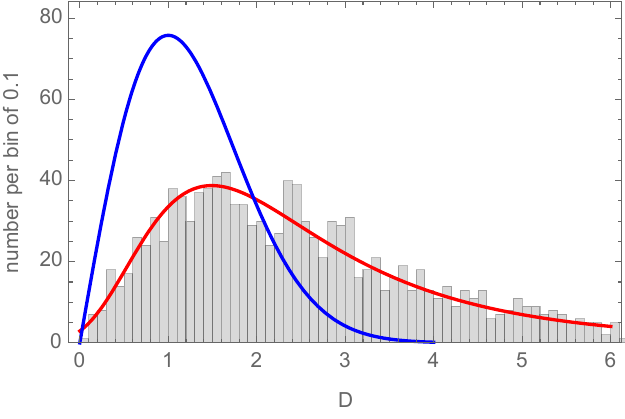}
\caption{Histogram of normalized position offset magnitudes $D$ (Eq. \ref{D.eq}) for the VLBI source 0016+731 with 1249 single-epoch measurements. The expected Rayleigh distribution with
scale 1 is shown as a blue curve. The best-fitting Fr\'{e}chet distribution with location $-0.40$, shape 1.20, and scale 1.84, is shown with a red curve.}
\label{frechet.fig}
\end{figure}

Fig. \ref{frechet.fig} illustrates the explicit inconsistency of the sample distributions of $D$ with the assumed normal distributions for large $n_{\rm obs}$. The histogram includes 1249 single-epoch position offsets for the ICRF3 source 0016+731. The blue curve is the expected Rayleigh[1] distribution. The red curve is the fitted Fr\'{e}chet distribution with location $-3.46$, shape 4.28, and scale 5.20. The negative location of the fit explains why the red curve is above zero at $D=0$, which is caused by a dearth of data points in close vicinity of the mean position. 

In addition to one-dimensional distributions of normalized offsets $D$, we explore the marginal distributions of standardized offset vectors $\boldsymbol{\delta}$ (Eq. \ref{st.eq}). The null hypothesis is that they have standard normal distributions ${\cal N}[0,1]$ in both components. This hypothesis is rejected with $p$-values of zero for all the 86 most frequently observed sources. We further fitted the best-matching statistical distributions from an extensive list of analytical bivariate PDFs with free fitting parameters. The best-fitting empirical distributions are predominantly of two types: Logistic and Student~$t$. The former have small mean parameters and scales mostly smaller than 1. The latter are found with scales slightly above 1 and degrees of freedom (dof) above 3. The Cauchy distribution is a particular case of Student~$t$ distribution with dof~$=1$, but our empirical distributions are less heavy-tailed. The standard normal distribution has dof~$=\infty$. For the test source 0016+731, the fit is Logistic$[0.35,1.56]$, which has an unusual high scale.

\subsection{Covariance and correlation} \label{sec:correl}

In the ICRF3 catalog, as well as in single-epoch VLBI position determinations, the {\it a priori} knowledge of the statistical uncertainties is represented by three numbers, namely, the formal errors of the celestial coordinates $\sigma_x$ and $\sigma_y$, and their correlation $\rho$. The errors, which are square roots of the corresponding variances, refer to the two fixed unit vectors $\hat{\boldsymbol{x}}$ and $\hat{\boldsymbol{y}}$ in the tangent plane at a chosen reference point $\boldsymbol{r}_{\rm ref}$ on the sky \citep[][Vol. 1]
{1997ESASP1200.....E}. We note that $\hat{\boldsymbol{x}}$, $\hat{\boldsymbol{y}}$, and $\boldsymbol{r}_{\rm ref}$ are three-dimensional vectors, whereas a positional offset $\boldsymbol{r}$ -$\boldsymbol{r}_{\rm ref}$ can be considered two-dimensional, as long as it is projected onto the tangent plane at the reference point. With the radial dimension folded in the traditional astrometric model, the covariance becomes a $2\times2$ matrix. These vectors and matrices are defined in a specific coordinate system, which is the ICRS for VLBI measurements, i.e., the equatorial coordinate system with a fixed vernal equinox. The vectors $\hat{\boldsymbol{x}}$ and $\hat{\boldsymbol{y}}$ are then the local directions toward east (increasing RA) and north (increasing declination), respectively.

The following analysis mostly concerns the distributions of position differences (or position offsets) $\boldsymbol{d}=\boldsymbol{r}-\boldsymbol{r}_{\rm ref}$ and the tuples $\{x,y\}$, which include the projections $x=\boldsymbol{d}\cdot \hat{\boldsymbol{x}}$ and $y=\boldsymbol{d}\cdot \hat{\boldsymbol{y}}$. Note that although the vector $\boldsymbol{d}$ is not orthogonal to $\boldsymbol{r}_{\rm ref}$, the incurred error is totally negligible in the small-angle approximation, being of the order of $\frac{1}{2}\,d^2$. These tuples can be treated as 2-vectors in the fixed tangential plane, and their distribution is therefore bivariate. By definition, the covariance matrix of $(x-x_0,y-y_0)^T$ is
\eb  
\boldsymbol C={\cal{E}}[(x-x_0,y-y_0)\,(x-x_0,y-y_0)^T],
\label{c.eq}
\ee 
where $\cal{E}$ is the expectation operator, and $\{x_0,y_0\}$ are the means of the corresponding variates. The covariance matrix takes the well known form:
\eb  
\boldsymbol C=\left(\begin{array}{cc}
   \sigma_1^2  &  \sigma_x\,\sigma_y\,\rho\\
    \sigma_x\,\sigma_y\,\rho & \sigma_2^2
\end{array}\right)
\label{c2.eq}
\ee  
only if the $(x,y)$ vectors are drawn from a binormally distributed population with standard deviations $\{\sigma_x,\sigma_y\}$ and correlation $\rho$. This implicit assumption is rarely spelled out in the astrometric literature, while having a crucial impact on the estimation of uncertainties, as we will see in the following.

\begin{figure}
    \includegraphics[width=0.47 \textwidth]{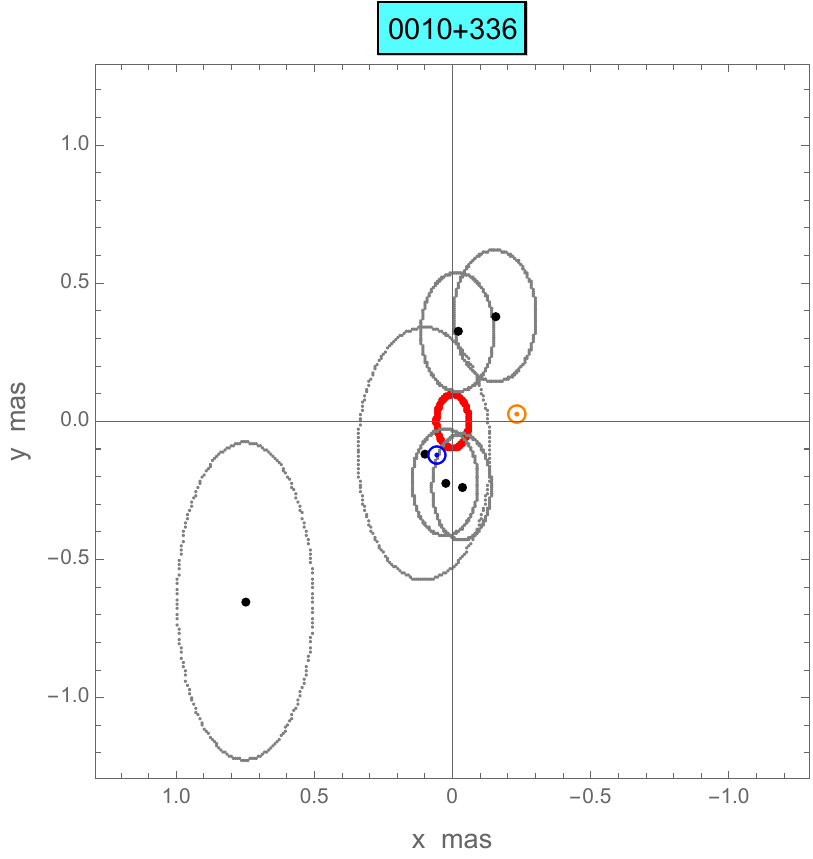}
    \caption{Single-epoch position measurements (black dots) and their associated error ellipses (dotted grey curves) obtained with the VLBI for the source IVS 0010+336. The plot is centered at the ML position computed in this analysis with the red ellipse representing its formal error ellipse. The blue circled dot marks the ICRF3 position for this source. The orange circled dot shows the position of the Gaia DR3 optical counterpart. Note the reversed $x$-axis, in accordance with the astronomical convention.}
    \label{0010.fig}
\end{figure}

The probability density function of the implicit binormal distribution is
\eb  
\begin{split}
{\rm PDF}[{\cal{B}}[\{x_0,y_0\}\,
\{\sigma_x,\sigma_y\},\rho]]=
\frac{1}{2\,\pi\,\sigma_x\,\sigma_y\,\sqrt{1-\rho^2}} \\
\exp{\left(-\frac{1}{2}\,(x-x_0,y-y_0)\,\boldsymbol B\,(x-x_0,y-y_0)^T\right)}.
\end{split}
\label{pdf.eq}
\ee  
Direct calculation using Eq. \ref{c.eq} shows that the normal matrix $\boldsymbol B=\boldsymbol{C}^{-1}$. Eq. \ref{c2.eq} follows directly from Eqs. \ref{c.eq} and \ref{pdf.eq}. In this paper, we investigate sets of single-epoch measurements obtained at various sessions with the VLBI for a collection of radio-loud quasars. Each single-epoch measurement comes with a covariance matrix in the tangential plane. This information captures the inherent uncertainties related to photon noise and the observational circumstances of the session, such as the orientation of the baselines, dispersion of multiple calibration parameters, etc. The range of associated standard deviations can be quite large, and the most distant outliers tend to have large formal uncertainties. The formal 2D uncertainty in the tangent plane can be well represented by the lines of constant probability, known for binormal distribution as error ellipses. Specifically, the quadratic form $(x-x_0,y-y_0)^T\,\boldsymbol C^{-1}\,(x-x_0,y-y_0)$
is distributed as $\chi^2$ with two degrees of freedom, and the corresponding confidence level at $\chi^2=1$ renders this equation of error ellipse:
\eb  
\frac{(x-x_0)^2}{\sigma_x^2}-2\rho\frac{(x-x_0)(y-y_0)}{\sigma_x \sigma_y}+\frac{(y-y_0)^2}{\sigma_y^2}=
1-\rho^2.
\ee
The error ellipse is always inscribed in a rectangle with sides $2\sigma_x$, $2\sigma_y$. In a special case when $\rho=1$, it degenerates into a straight line. We note that unlike the 1D normal distribution, the probability that a random position measurement falls within an error ellipse is only 0.393. Fig. \ref{0010.fig} shows an example of the observed configuration of position measurements (black dots) and their formal error ellipse (grey dotted curves) for the source IVS 0010+336 with six available single-epoch measurements. The error ellipses are conspicuously elongated in the south-north direction indicating a greater uncertainty of astrometric measurements of declination.

\subsection{Estimation of the mean position}
\label{ml.sec}
Quasars are generally assumed to be stationary astrometric sources with fixed positions on the sky.\footnote{Evidence has recently emerged that some quasars are not stationary, however \citep{2020RNAAS...4..108T,2023AJ....165...69T}.} The observed dispersion of single-epoch positions is then ascribed to random observational errors. Therefore, the consistent way of deriving the sample-mean position within the implicit hypothesis of binormality is to use the maximum likelihood (ML) principle, which defines the mean as the point of the highest joint probability. In the absence of evidence about correlations between individual measurements, we have to assume that they are independent, and the joint PDF is the product of individual probabilities (Eq. \ref{pdf.eq}). The log-likelihood function includes the sum of quadratic forms
$(x_i-\bar x,y_i-\bar y)\,\boldsymbol C_i^{-1}\,(x_i-\bar x,y_i-\bar y)^T$, which has maximum at the most probable position $(\bar x,\bar y)$. This optimization problem has a unique solution at
\eb  
(\bar x,\bar y)=\left(\sum_i \boldsymbol C_i^{-1}\right)^{-1}\; \sum_i \boldsymbol C_i^{-1}\,(x_i,y_i)^T.
\label{ml.eq}
\ee  
The leading factor is the covariance matrix of the ML position. The tangential coordinates entering this equation refer to a common origin at the celestial position $\boldsymbol r_{\rm ref}$. The computation is invariant to the choice of reference position as long as it is sufficiently close to the single-epoch positions.

In Fig. \ref{0010.fig}, the computed ML position is at the origin of the local coordinate system, because we use this position as a new reference point. The red curve represents the formal error ellipse ($\chi^2=1$) of that position. The significant elongation in the declination dimension is inherited from the predominant direction of the single-measurement ellipses. The ML error ellipse is significantly smaller than each individual ellipse, because in this case, a few measurements with approximately equal uncertainties provide substantial reduction of formal dispersion. The weights in Eq. \ref{ml.eq} being effectively quadratic, the ML estimation is most sensitive to data points with smaller formal variances. It is therefore crucially important to see if the formal errors faithfully capture the empirical dispersion, and, more generally, if the observed sample distributions are consistent with the implicit binormal distribution. The plot also illustrates the importance of using the full covariance (rather than just $\sigma_x$ and $\sigma_y$) when the mean VLBI positions are compared with other astrometric data, such as the optical Gaia positions.

\subsection{Standardization of position measurements}
\label{sta.sec}
For a given source, we are investigating a sample of position measurements with widely different formal covariances. The derived ML position is consistent with the implicit statistical hypothesis and becomes our new point of reference, but does the observed sample distribution confirm this initial assumption? To answer this question, a couple of standardized variables $\{u_{x(i)},u_{y(i)}\}$ is introduced, such that
\eb  
\boldsymbol{\delta}_i\equiv (u_{x(i)},u_{y(i)})^T =
\boldsymbol C_i^{-1/2}\,(x_i-\bar x,y_i-\bar y)^T.
\label{st.eq}
\ee  
Note that the powers of $\boldsymbol C$ throughout this paper are {\it matrix powers}, not by-element operations. With this substitution of variables, the assumed PDF of a single measurement in Eq. \ref{pdf.eq} simplifies to
\eb  
{\rm PDF}[{\cal{B}}[\{\bar x,\bar y\}\,
\{1,1\},0]]=
\frac{1}{2\,\pi} \\
\exp{\left(-\frac{1}{2}\,\boldsymbol{\delta}_i^T\,\boldsymbol{\delta}_i\right)}.
\label{pdfsta.eq}
\ee  
The standardized formal distribution is rotationally symmetric with both standard deviations equal to unity and correlation equal to zero. Thus standardized single-epoch measurements, within the underlying hypothesis, become statistically uniform, in that they are expected to be drawn from the same population with a standard binormal distribution.

\section{Analysis}  \label{sec:analysis}

The fundamental goal of this work is to investigate the variability of source positions. 
For simplicity and clarity, we use offsets or differences from standard reference points -- the official ICRF3 X/S positions -- rather than directly using absolute Right Ascension and Declination coordinates.  
Total magnitudes of offsets (typically denoted ``total'' herein) are given along with their orthogonal R.A.\ and decl.\ components.  

We consider several metrics of variability to characterize the measured source position differences. 
As there are several forms and nuances of variability in a given dataset, we differentiate between three broad categories of variability for our time series in this work.  The first two are roughly analogous to accuracy and precision, statistical measures of typical offsets and the dispersion in the overall measurements for a given source.  The third category explored here is that of coherent trends of apparent motion or drift over time, which is also of great importance for assessing the quality of reference frame sources that might otherwise appear `stationary' based on overall statistical population measures. 

Standard statistics such as the mean, median, weighted root mean square (wrms), and standard deviation are intuitive and widely used, though they can be somewhat limited owing to outlier sensitivity and formal assumptions about the data being independent and identically (specifically, Normally) distributed.  
Descriptions of our statistics are given below, followed by a comparison and notable points of consideration.

\subsection{Typical Offsets -- Jitter} \label{sec:jitterOffsets}

\begin{figure}
\includegraphics[width=\linewidth]{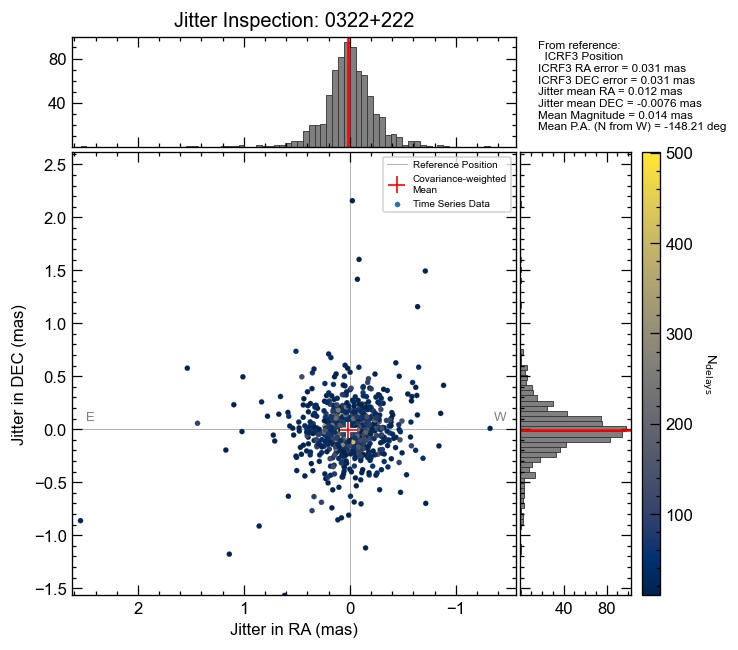}
\caption{Example measurement ``jitter'' offsets from the ICRF3 catalog position for a single source, 0322+222. 
The color of the scatter points corresponds to the number of delays observed for this source in each individual diurnal session measurement. 
Marginal distributions of the R.A. and Decl. component offsets are shown on the top and right sub-axes, respectively.  
The covariance-weighted mean value and its components are indicated in the top corner.   
}
\label{fig:jitter_example_source}
\end{figure}

\begin{figure}
\centering
\includegraphics[width=\linewidth]{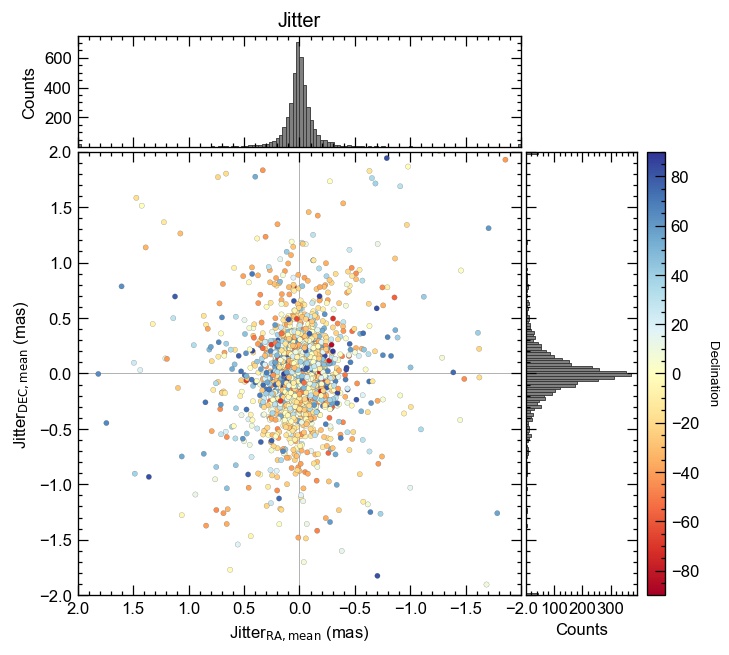}
\includegraphics[width=\linewidth]{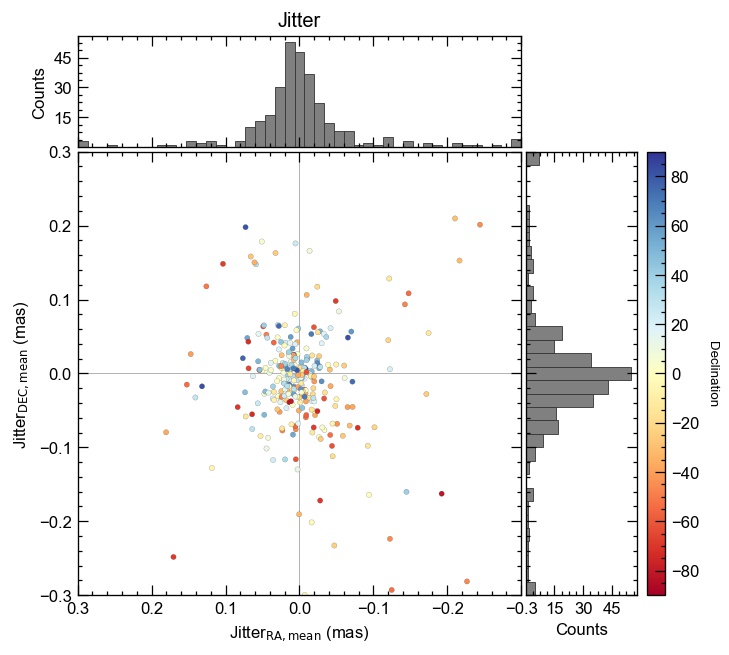}
\caption{Jitter -- covariance-weighted mean offsets from ICRF3 positions, for time series of all ICRF3 sources (top) and the defining sources (bottom). 
The data points are colored by source declination.
Marginal distributions of the R.A. and Decl. components are displayed in the side axes, and show a slightly larger dispersion in the N-S than the E-W direction owing to typical baseline geometry.
}
\label{fig:jitter_allsrc_scatter}
\end{figure}

\begin{figure}
\includegraphics[width=\linewidth]{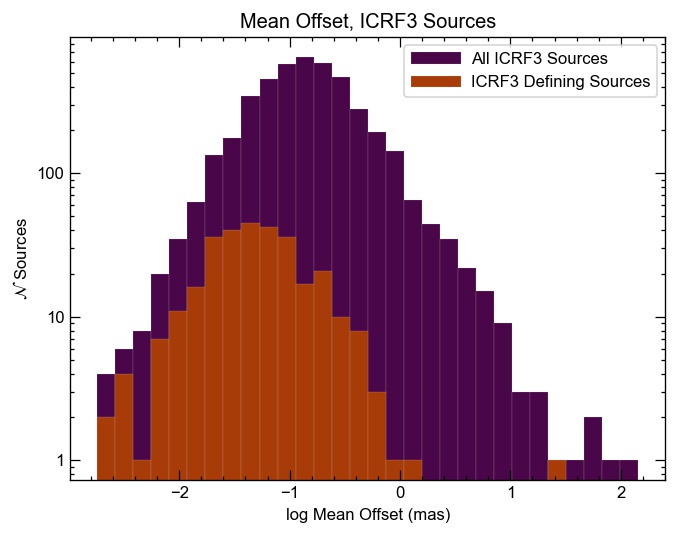}
\caption{Distributions of the total magnitude (combined R.A. and Decl.) of jitter (covariance-weighted mean) offsets for the full ICRF3 source list and the ICRF3 defining source list.  }
\label{fig:jitter_histograms_def_full}
\end{figure}

The typical offset from the ICRF3 coordinates is one crucial estimate of a source's variability that can readily highlight its suitability for use in reference frames. 
While all measurement distributions will have some dispersion, it is critical that on average the values center as closely as possible to the expected reference point. 
Outliers, correlations, mis-characterizations of the true uncertainties, and similar aspects of the data can greatly impact the effort to accurately summarize the distribution's properties, however. 
The measured sky coordinate components are not independent, therefore statistics that inherently treat right ascension and declination as paired values are more suitable than more rudimentary 1D measures such as the simple mean or median that treat each component separately.

We dub the collected set of individual measured offsets for a particular source over time the ``jitter'', and from the jitter we calculate a variety of values, including the typical offset. 
Figure~\ref{fig:jitter_example_source} shows an example of these jitter values for a single ICRF3 source, 0322+222. 
Various statistics capable of pairing values were explored in order to describe the typical jitter offset, such as the weighted geometric median and mean, but the covariance-weighted mean is the most straightforward statistic that fully utilizes the session coordinate covariance matrices.  
The details of calculating a covariance-weighted mean are discussed further in Sec.~\ref{ml.sec}. 
Figure~\ref{fig:jitter_allsrc_scatter} shows the calculated covariance-weighted mean offsets for each source as individual datapoints.  Zero here corresponds to no offset from a source's official ICRF3 position.

For summary statistics over all sources, we apply a reliability filter to our calculations, only considering sources here with at least four observing sessions, at least 10 delays measured in each session, and observations spanning at least two years. 
Figure~\ref{fig:jitter_histograms_def_full} shows the distributions of the magnitudes of the covariance-weighted mean offsets (radius from reference position) for the sources. 
The ICRF3 defining sources have a median jitter magnitude offset of 48~$\mu$as, whereas including the densifying sources, the median value is 137~$\mu$as.
From this perspective, the defining sources are on average more stable than the densifying sources, though that is at least partly owing to the defining sources typically having much longer series of observations than the densifying sources.

\subsection{Dispersion -- wrms, $Q68$, and Excess Uncertainty} \label{sec:ExcessError}

\begin{figure}
\includegraphics[width=\linewidth]{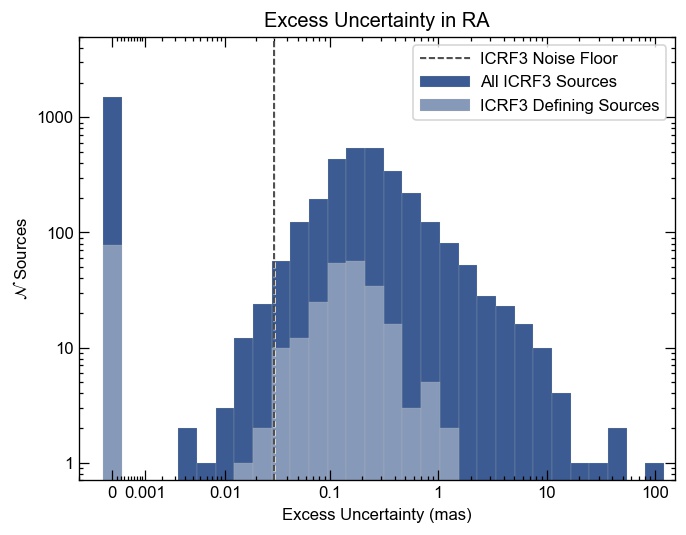}
\includegraphics[width=\linewidth]{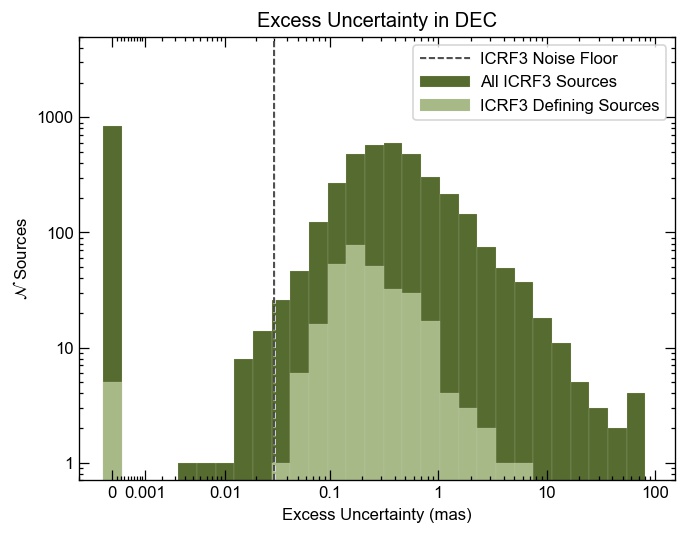}
\caption{Histograms of excess error in R.A. and Decl. components computed from the time series of each source.  
The full ICRF3 source list is denoted by the darker shading while the defining sources are denoted by the lighter shading.  
The ICRF3 X/S noise floor values of 30~$\mu$as are shown by the vertical dashed lines, for reference. 
}
\label{fig:excessvariance_histograms_def_full}
\end{figure}

\begin{figure*}
\centering
\includegraphics[width=0.48\linewidth]{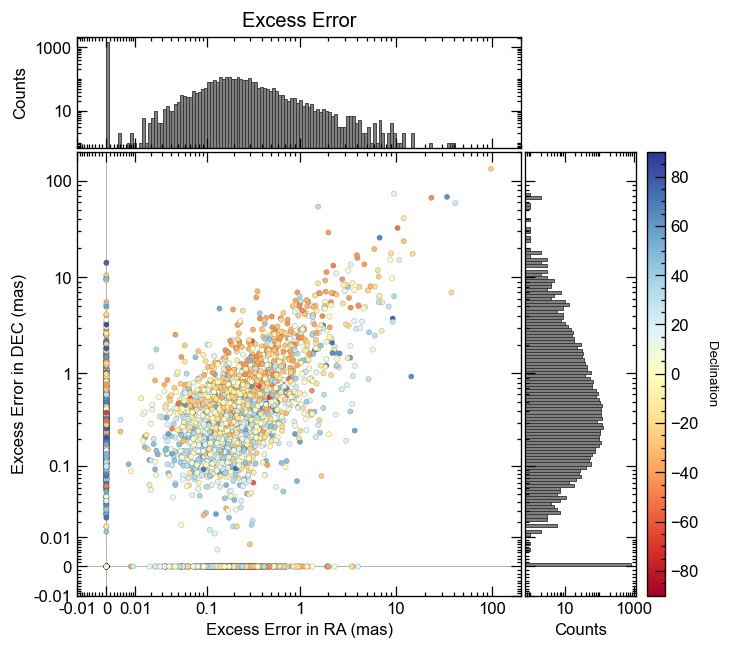}
\hspace{0.5cm}
\includegraphics[width=0.48\linewidth]{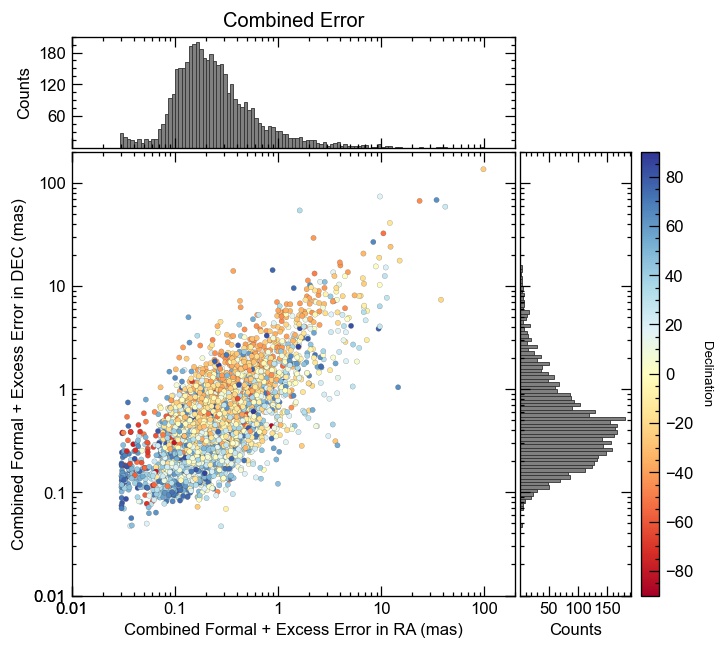}
\includegraphics[width=0.48\linewidth]{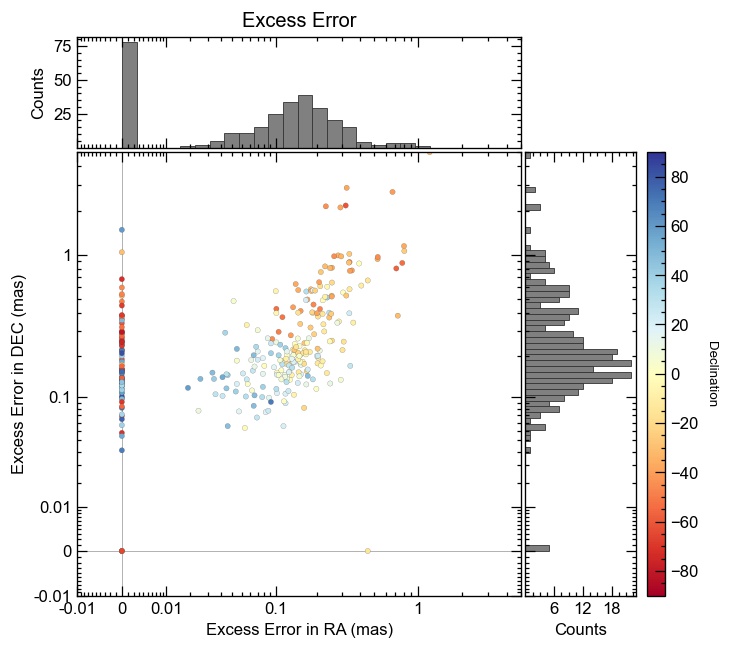}
\hspace{0.5cm}
\includegraphics[width=0.48\linewidth]{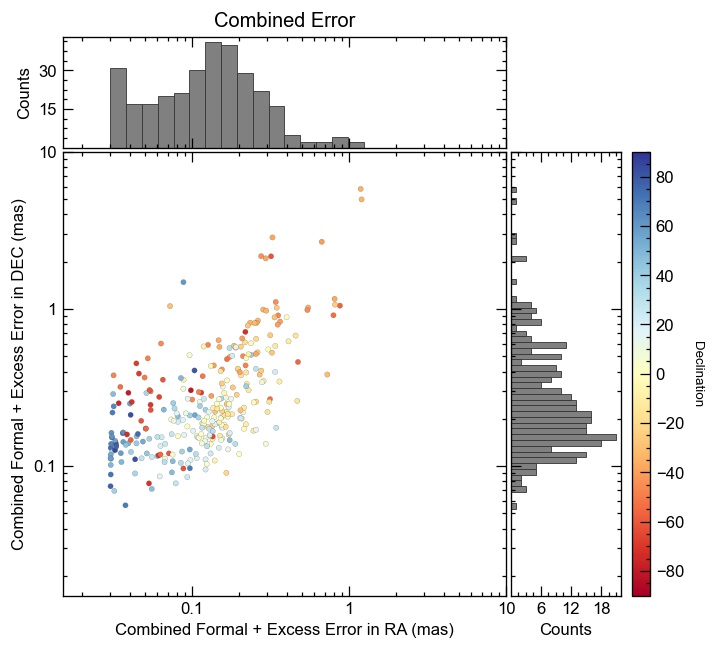}
\caption{Comparison of R.A. and Decl. component values of excess error.  
The left column shows the excess errors alone, while the right column shows combined formal + excess error values.
The top row includes all ICRF3 X/S sources and the bottom row shows only the ICRF3 X/S defining sources. 
The data points are colored by source declination, and show that excess and combined errors are somewhat larger for sources observed at lower declinations, though with a large spread.
Marginal distributions of the R.A. and Decl. components are displayed in grey on the side axes.
}
\label{fig:excesserror_allsrc_scatter}
\end{figure*}

The dispersion of a dataset is another fundamental aspect of its variability, and there are numerous ways to characterize this. 
The closely related standard deviation and weighted root mean square (wrms) are two classic metrics under the assumption of Normally-distributed data.  
Fits to such data should satisfy $\chi^2_\nu$ = 1 when the uncertainties perfectly reflect the distribution of measured data.  
The source position fits from global solutions are complicated however, and moreover the ICRF3 X/S sources have an imposed noise floor of 30 $\mu$as in R.A. and decl., where any solution uncertainties below this level are inflated to maintain this minimum \citep{2020A&A...644A.159C}. 
Frequently, the scatter seen contained within the time series can exceed the reported ICRF3 uncertainties.  We introduce here an ``excess variance'' to be added to the source coordinate uncertainties such that the combined variances (formal plus excess) on computed position offsets yield $\chi^2_\nu$ = 1, in order to capture any extra or unmodeled variance and make it consistent with the formal assumptions. 

For reference positions, we use the typical offset seen in the time series -- the covariance-weighted mean jitter position described above, computed utilizing the full coordinate covariance matrices for each session. 
Then, for each coordinate component, 1 $\mu$as of white noise is incrementally added in quadrature to the formal (inflated) time series uncertainties, until the resulting time series of position offsets result in $\chi^2_\nu$ = 1.  This excess error can be combined with the formal error in quadrature to give a better representation of the true statistical error in the data. 

Histograms of these excess uncertainties are shown in Figure~\ref{fig:excessvariance_histograms_def_full}, and comparisons of excess errors in each component are plotted in Figure~\ref{fig:excesserror_allsrc_scatter}. The data there are colored by source declination, and show that more southerly sources tend to have higher excess errors than northern-sky sources, especially in the declination component; this will be discussed in more detail in Sec.~\ref{sec:byDEC}.  Many sources yield zero excess error -- meaning the formal error adequately represents the data under the formal assumptions -- however most sources require some excess error, and on average they require similar amounts in both coordinate components as shown by the positive correlations in the plots. 

We characterize the statistical spread in the source time series using the wrms as well as the astrometric ``noisiness'' metric $Q68$ introduced in Sec.~\ref{test.sec}.  The combined error, comprising both the formal and calculated excess, is used to ensure the unmodeled error not captured in the formal errors is included.

\subsection{STR: Smoothed Time-series Range}
\label{sec:STR}

Several measures of variability have been discussed to characterize the typical offsets and dispersion of the observed positions of sources, but these do not necessarily capture long-term coherent variability.
It is possible for source measurements to have a relatively tight dispersion in values estimated during similar epochs and still have notable coherent trends in the data whose significance may become diluted by large number statistics. 
Many sources show somewhat flat overall trends (with noise) for long periods but with occasional obvious deviations that may only span a few percent of the data, and therefore get somewhat washed out by statistical methods. 
Even spurious cases such as the large jump seen for 3C48 could even be muted by statistical methods, depending on how many epochs are sampled.
Therefore it is important to investigate coherent trends in the data, as well as the statistical offset and dispersion measures discussed above.
For reference frame work, where position stability is of the utmost importance, it is desirable to determine if a source exhibits significant variability regardless of its cause. 
We have an intuitive and robust approach to quantify this in a simple manner, described below.

First, we smooth the time series of each source to reduce scatter, and to help reveal any underlying trends. 
Rather than standard methods that use a constant number of data points for filtering, which has the effect of smoothing trends over potentially quite different time scales, we instead filter the time series using a specified \textit{time} window.  
The choice of width or timespan for the filter window is important, as short windows will reduce the scatter less and longer windows could smooth over trends on time periods of interest.  
Several windows were tested, ranging from 30 to 360 days, and a 120-day window was found to have the optimal balance overall for reducing scatter as well as longer windows, but without blending obvious features and still being sensitive enough for months-long trends. 
For the occasional epochs when there are gaps or too few data points within a particular rolling 4-month window for the filter to be effective, continuity and smoothness were maintained by using a nominal window of N=3\% of the array length in those periods instead. 
This yielded good results without jumps or discontinuities due to sampling. 

The choice of filtering statistic can also impact the quality of smoothing.  
Several variations of means and medians were tested, both in terms of the flavor (such as standard mean/median vs geometric mean/median), and in terms of the weighting (standard weighting by each component independently as a control vs covariance-weighted versions of mean/median).
The best results overall were achieved by using the covariance-weighted mean for the filter, for two primary reasons: it utilizes the full formal covariance matrix of each session; and while the median is less biased by outliers than the mean for large datasets, the sometimes sparse sampling within certain time windows results in more small jumps than the smoother results of the (covariance weighted) mean.

\begin{figure}
\includegraphics[width=\linewidth]{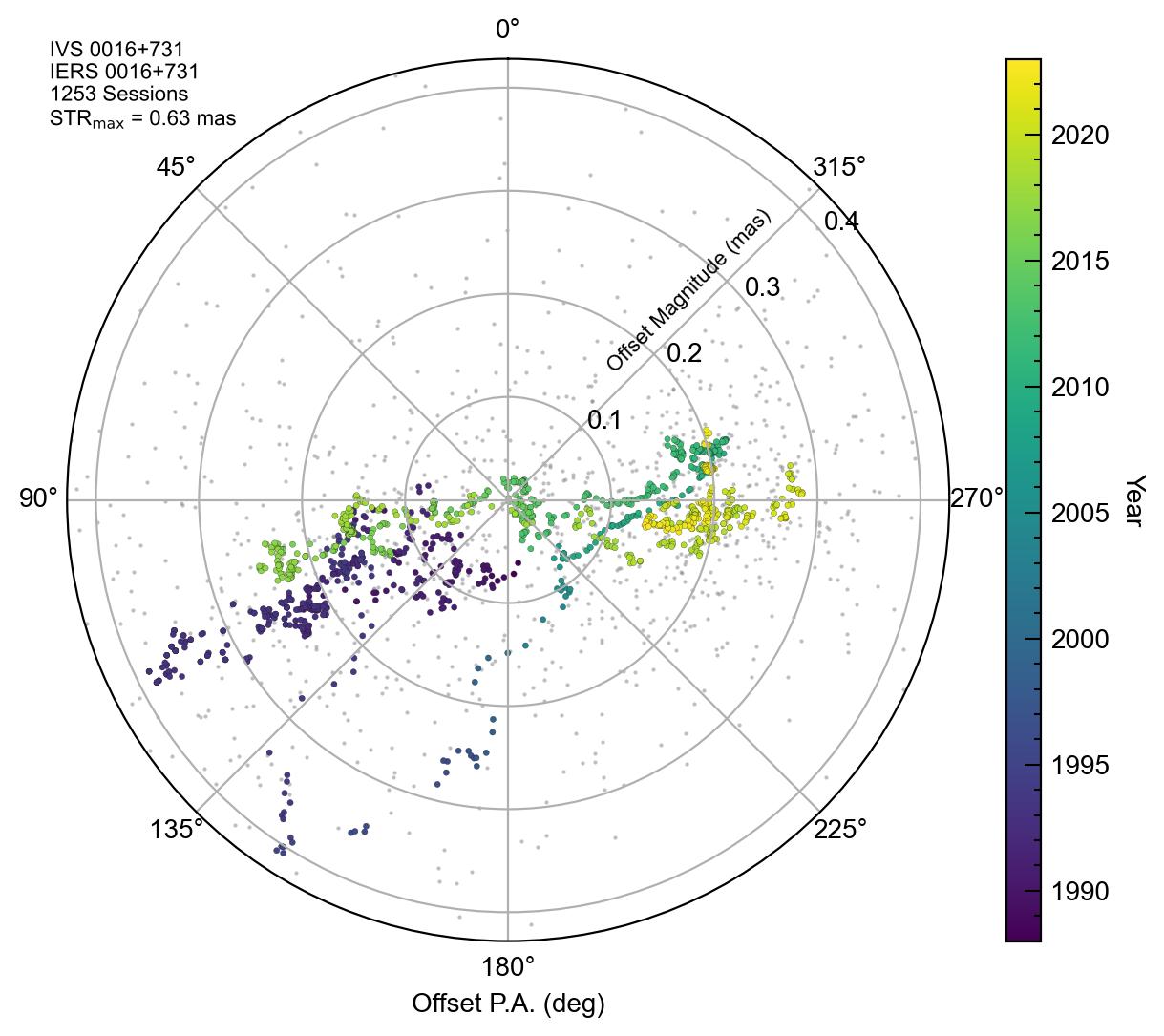}
\caption{ Smoothed time series positions on the sky for ICRF3 defining source 0016+731. Time is denoted by the color scale. Gray background points are the unsmoothed time series.  This reveals clear coherent position variability on the order of $\sim$0.4 mas on several-year timescales. 
This figure is available as an 8-second animation in the online version of this article.  In the animation, a scroll bar progresses smoothly through the years spanning the astrometric VLBI sessions, and the scatter points representing the single-epoch position estimates appear at the time of their observations, tracking the observed positions over time as they appear on the sky, ultimately ending with the static image where all observations are included. }
\label{fig:STR_example_2D}
\end{figure}

\begin{figure*}
\includegraphics[width=\linewidth]{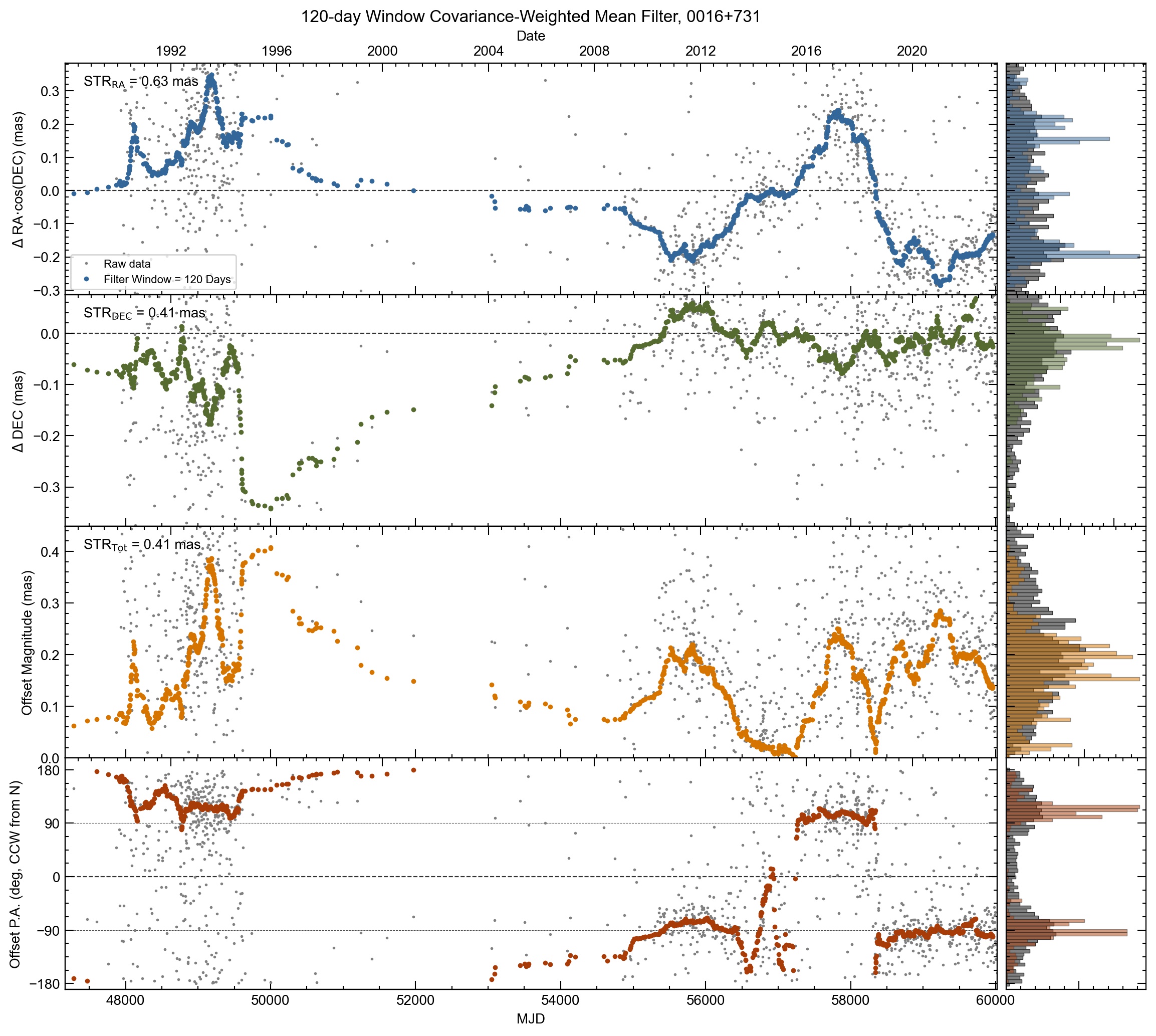}
\caption{ Smoothed time series and component STR metrics for ICRF3 defining source 0016+731. Time is on the x-axis. Gray background points are the unsmoothed time series, while colored points show the smoothed series.  Vertical axis ranges have been optimized for the smoothed series.  Panels from top to bottom show the RA$\cdot$cos(DEC) and DEC offset components, the total offset magnitude, and the offset position angle (CCW from North).   }
\label{fig:STR_example_4panel}
\end{figure*}

The total peak-to-peak range of observed offsets is a simple yet instructive measure of the total variability.
While this would have been useless for the raw data, owing to the numerous extreme outliers, the peak-to-peak range of smoothed offsets is rather a useful measure, and intuitively corresponds to features that can be readily identified. 
We refer to this measure as STR (smoothed time-series range) hereafter, for compactness. 
Again we restrict these calculations to sources having at least 4 sessions, at least 10 observed delays per session, and observations that span at least two years.
Figure~\ref{fig:STR_example_2D} shows an example of the smoothed time series for ICRF3 defining source 0016+731 plotted as offset positions on the sky, and Figure~\ref{fig:STR_example_4panel} shows the same time series as well as STR values, broken down by coordinate component.

Collating the STR results from all ICRF3 sources, 541 sources have a maximum STR of less than 0.03~mas, 215 sources are better than 0.01~mas, and 179 are better than 0.005 mas.  
On the high variability end, 698, 81, and 27 sources have maximum STR $>$ 1.0~mas, 5.0~mas, and 10.0~mas, respectively.  
3C48 has the highest STR value of 57.8 mas, corresponding to the sudden `jump' observed circa 2018.
For the defining sources, 38 have a maximum STR of $<$ 0.5~mas, 18 sources are $<$ 0.4~mas, and only 5 sources are better than 0.3~mas. 
Many of the defining sources also have quite high maximum STR values -- 138, 18, and 7 sources have maximum STR values $>$ 1.0~mas, 5.0~mas, and 10.0~mas, respectively.
This is at least partly due to the much longer observational histories of the defining sources than the majority of the densifying sources.

\begin{figure*}
\includegraphics[width=0.465\linewidth]{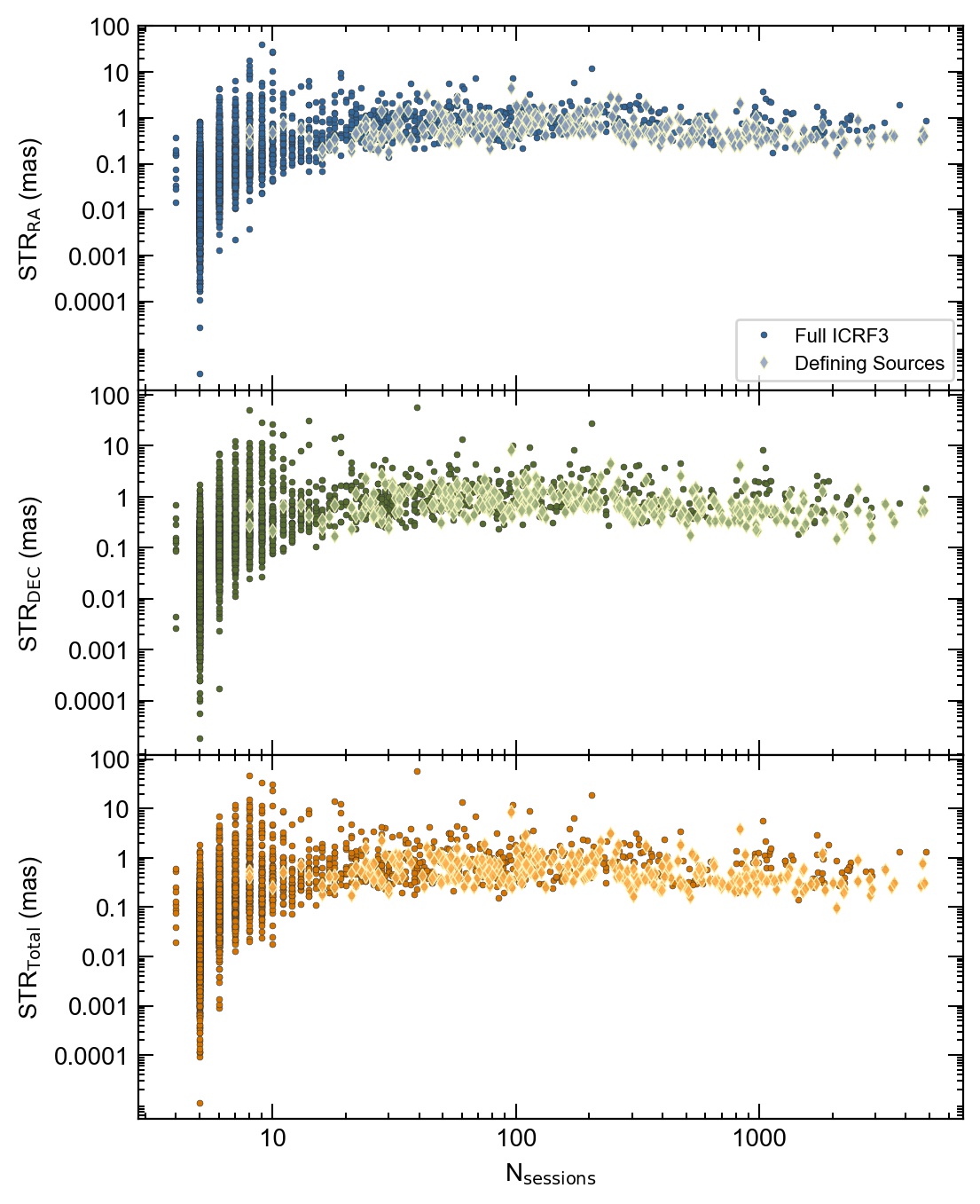}
\hfill
\includegraphics[width=0.45\linewidth]{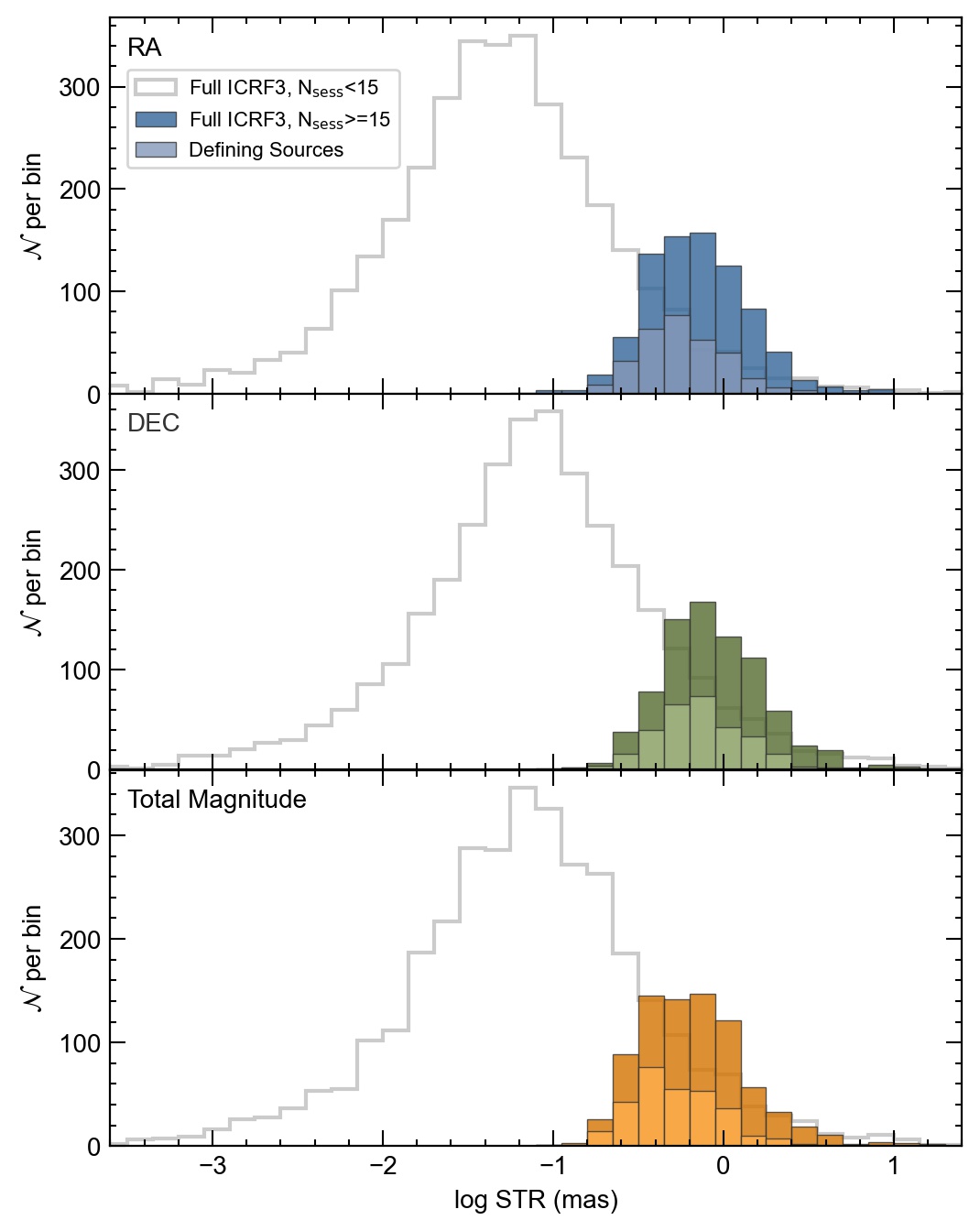}
\caption{ 
Left: Source STR (smoothed time-series range) values as a function of the number of sessions in their time series, separated by coordinate component.  From top to bottom are right ascension, declination, and total magnitude.  Sources with fewer sessions have seemingly low average STR values with large scatter, dramatically converging on a stable flat trend at a threshold of $\sim$15~sessions.  
Right: Histograms of the STR distributions for all ICRF3 sources, again separated by component.  The defining sources are plotted with lighter shading, and the full set of ICRF3 sources with at least 15 observing sessions appear in darker shading. The ICRF3 sources with fewer than 15 sessions are also shown for reference with the light gray line, and demonstrate that time series with few observations ($\leq$15 sessions) are distributed differently.
}
\label{fig:STR_hists}
\end{figure*}

From this perspective, all quasars are astrometrically variable over months to years timescales, relevant to reference frame solutions. 
Figure~\ref{fig:STR_hists} show histograms of the STR values by component for ICRF3 sources. 
The full ICRF3 source list is clearly (at least) bimodal in each of the R.A. and decl. components, with a subset of sources having typical STR variability $\lesssim$ 0.1~mas, and another subset centered around 1.0~mas.  
There are likely to be multiple smaller subsets superposed, given the broad tails.  
The defining sources are closer to a single-modal distribution with peak around 0.5~mas -- on the high end compared to many of the densifying sources, though this is because of the typically much longer observational history.

\pagebreak

\section{Results and Discussion} \label{sec:results}

\begin{deluxetable*}{lD|DD|D} 
\caption{Overall Statistics for VLBI Measurements of ICRF3 Sources}
\tablehead{ \colhead{Statistic} & \multicolumn{4}{c}{X/S Defining Source Set} & \multicolumn{4}{c}{X/S Full ICRF3 Set} \\
                                     & \multicolumn2c{Mean} & \multicolumn2c{Median} & \multicolumn2c{Mean} & \multicolumn2c{Median} }
\decimals
\startdata
\multicolumn{9}{c}{Typical offset -- jitter} \\
jitter$_\mathrm{RA}$ ($\mu$as)                       &       4.9   &    8.8     &      22.4  &   59.5   \\
jitter$_\mathrm{DEC}$ ($\mu$as)                      &      -2.8   &   -2.3     &       9.4  &    3.2   \\
jitter$_\mathrm{Tot}$ ($\mu$as)                      &       5.7   &    9.1     &      24.3  &   59.6   \\
jitter$_\mathrm{Tot}$ Position Angle (deg, E from N) &     120.0   &  104.6     &      67.3  &   86.9   \\
\hline
\multicolumn{9}{c}{Dispersion -- wrms, excess error, $Q68$} \\
wrms$_\mathrm{RA}$  ($\mu$as)                        &     348.5   &  256.1     &     567.9  &  272.2   \\
wrms$_\mathrm{DEC}$ ($\mu$as)                        &     473.4   &  334.5     &     927.8  &  442.9   \\
wrms$_\mathrm{Tot}$ ($\mu$as)                        &    1024.7   &  720.2     &    1033.2  &  437.2   \\
$s_\mathrm{excess,RA}$ ($\mu$as)                     &     138.8   &  115.0     &     351.2  &  123.0   \\
$s_\mathrm{excess,DEC}$ ($\mu$as)                    &     353.7   &  202.0     &     743.3  &  272.5   \\
$s_\mathrm{excess,Tot}$ ($\mu$as)                    &     394.8   &  243.8     &     892.9  &  348.1   \\
$s_\mathrm{combined,RA}$ ($\mu$as)                   &     165.0   &  130.0     &     477.7  &  208.4   \\
$s_\mathrm{combined,DEC}$ ($\mu$as)                  &     373.1   &  219.3     &     917.3  &  402.5   \\
$s_\mathrm{combined,Tot}$ ($\mu$as)                  &     419.1   &  264.8     &    1075.5  &  476.2   \\
$Q68$                                             &      1.93   &    1.91    &     2.05   &   1.86   \\
\hline
\multicolumn{9}{c}{Coherent trends in time -- STR} \\
STR$_\mathrm{RA}$  ($\mu$as)                         &     683.0   &   541.6    &     923.1  &  672.4   \\
STR$_\mathrm{DEC}$  ($\mu$as)                        &     884.2   &   676.6    &    1279.6  &  804.3   \\
STR$_\mathrm{Tot}$  ($\mu$as)                        &     662.3   &   485.0    &    1039.3  &  625.2   \\
\enddata
\tablecomments{
Mean and median statistics for the overall distributions of the VLBI position variability metrics, for the full ICRF3 sample as well as the ICRF3 defining source sample.
Source jitter statistics are reported for individual coordinate components, as well as for the total magnitude (RA and DEC combined) and the position angle of the mean/median total magnitude. 
The jitter statistics presented here are the covariance-weighted mean and covariance-weighted geometric median of the set of typical jitter offset values for each source.  
The position angles of the total offset follow the customary astronomical definition, in degrees East from North. 
$s_\mathrm{excess,RA}$ and $s_\mathrm{excess,DEC}$ denote excess variance in R.A.\ and decl., respectively, and $s_\mathrm{excess,Tot}$ is the quadratically-summed excess error in both directions.  
$s_\mathrm{combined}$ signifies the excess and formal errors combined. 
The weighted root mean square value of each component is denoted by the usual wrms.
$Q68$ refers to the 0.68-quantile of the normalized position offsets, a measure of the ``noisiness'' of the distributions (see Sec.~\ref{test.sec}).
STR refers to the smoothed time series range, and quantifies the maximum range in the offsets over time, after smoothing with a rolling four-month time window (see Sec.~\ref{sec:STR}).
}
\label{tab:jitter1}
\end{deluxetable*}

\subsection{Variability Metric Comparison, and Ranking Sources} \label{sec:VariabilityComparison}

\begin{figure*}
\includegraphics[width=0.48\linewidth]{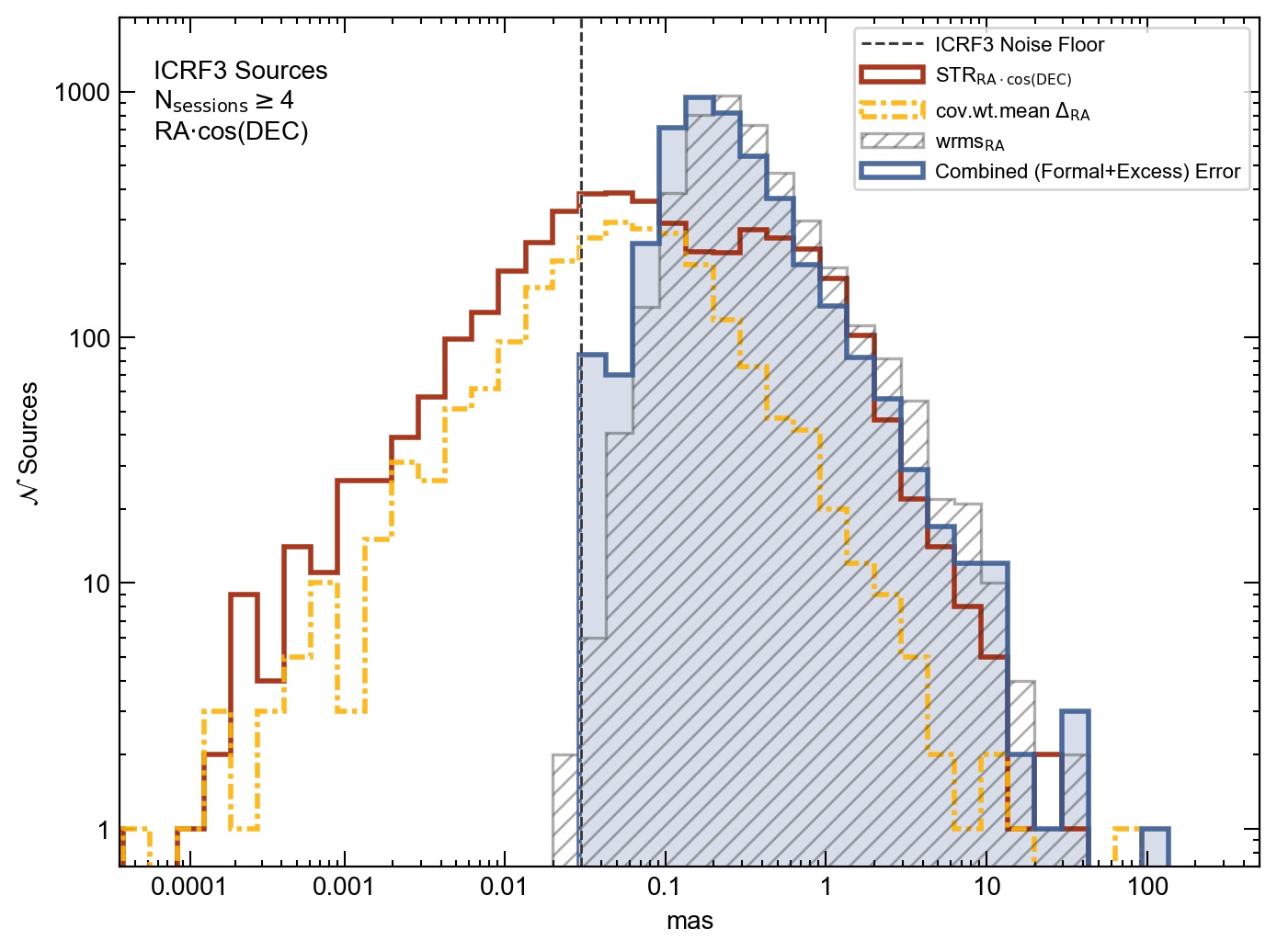} 
\includegraphics[width=0.48\linewidth]{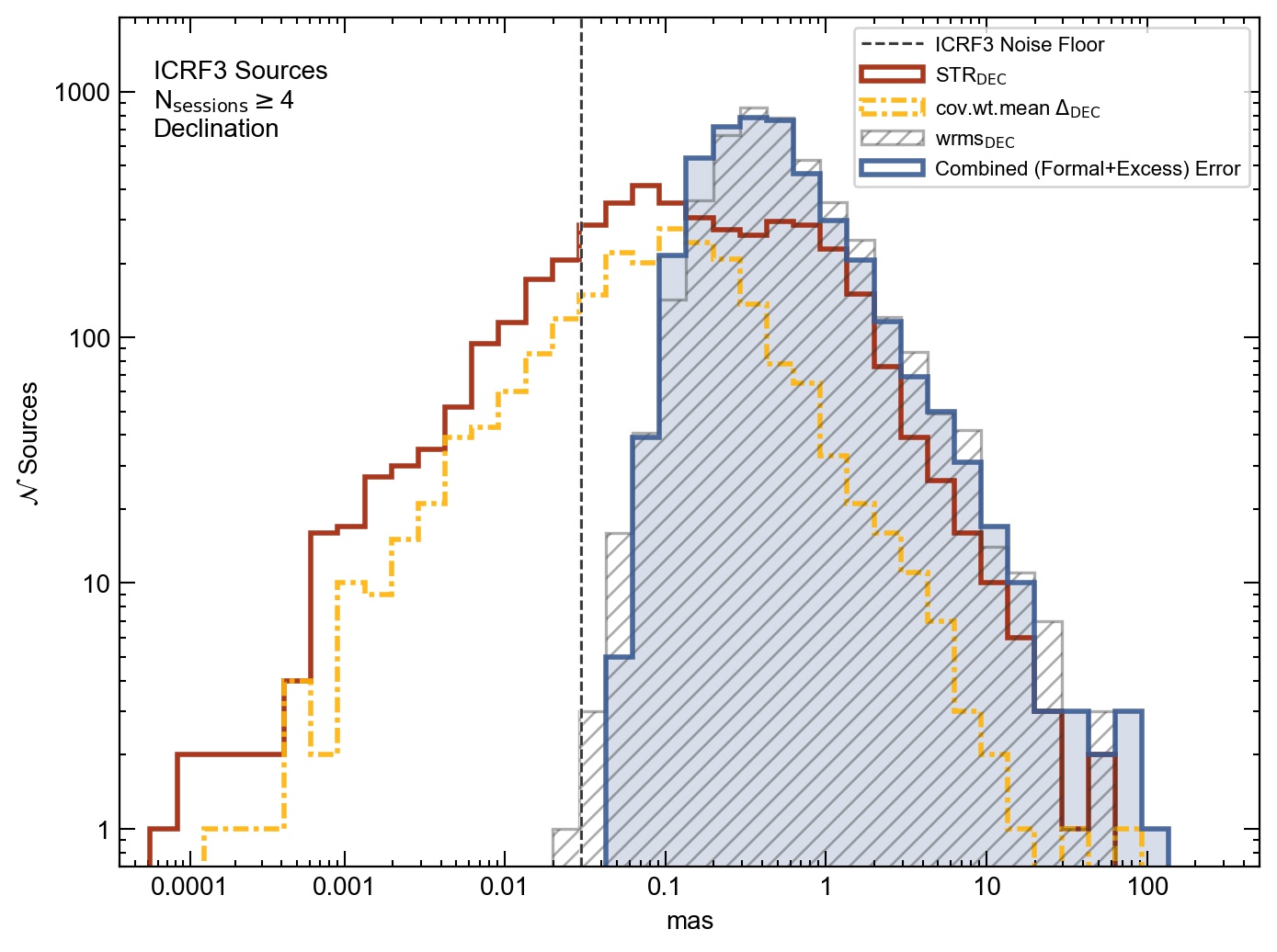} //
\includegraphics[width=0.48\linewidth]{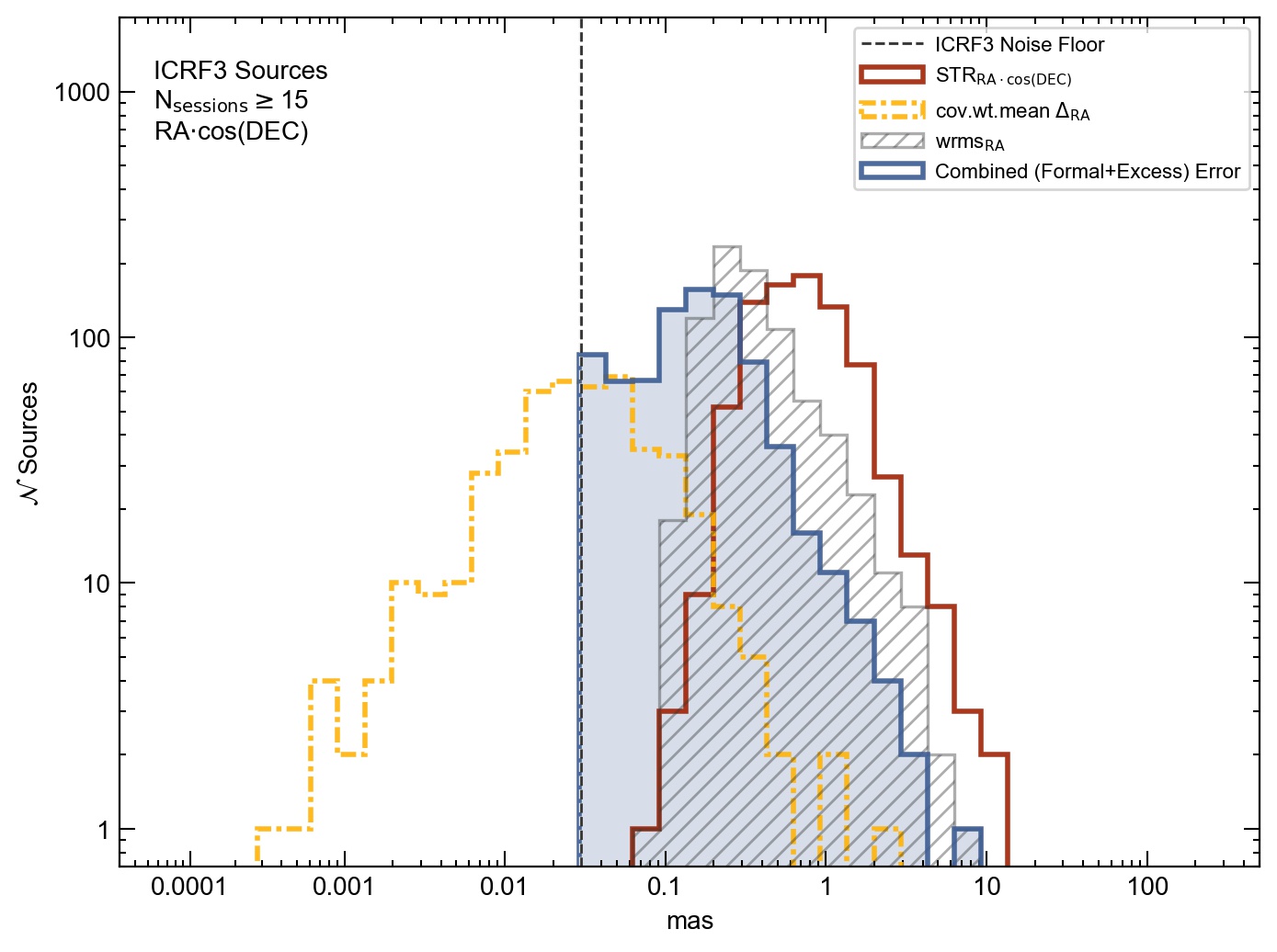} 
\includegraphics[width=0.48\linewidth]{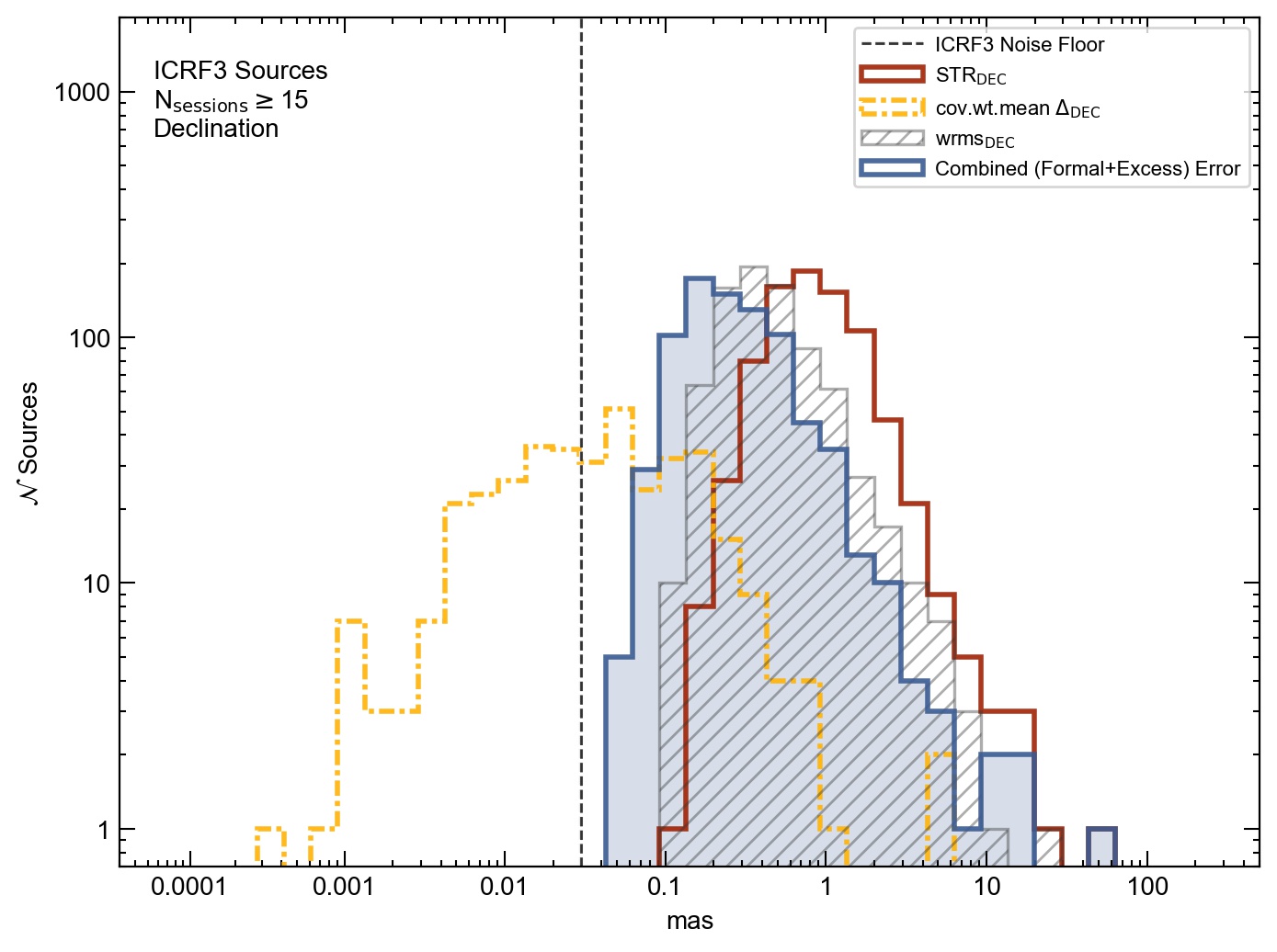} 
\caption{Distributions of variability metrics in milliarcseconds for the full ICRF3 source list, by tangential coordinate component: right ascension in the left panels, declination in the right panels. The top row shows the distributions for sources detected in at least four diurnal sessions (and at least 10 delays in each), reflecting nearly all sources. The bottom row shows the same distributions but only for sources with 15 or more sessions, which excludes a large fraction of sources but represents those with more robust values. 
The dashed vertical line denotes the 30~$\mu$as ICRF3 noise floor in each component, for reference. The red steps denote STR (smoothed time-series range) values, the yellow steps show the distribution of covariance-weighted mean jitter values, the grey hatched steps show the wrms, and the blue shaded histogram shows the combined (formal plus excess) error. 
}
\label{fig:excessvariance_all_full}
\end{figure*}

The celestial sources used in astrometric and geodetic observations that create the foundation of our reference frames can be ranked into various groups of quality, based on estimates of their stability.  Given the rich and complex nature of the observed time series, sources can obviously be characterized as more or less variable depending on which metric is used.  Not only is it important to consider multiple statistics in assessing the variability of sources, it is essential to look at additional metrics beyond the standard population statistics, and moreover it is critical to consider the \textit{combination} of multiple types of variability in order to capture the multiple ways in which measured positions can appear to change over time. 

For an example highlighting the value of these additional metrics of variability, and the importance of using different types of variability, consider the frequently observed ICRF3 source 2229+695.  It has a total (R.A. plus decl.) formal error of 43 $\mu$as, total wrms of 191 $\mu$as, and even typical jitter offset from its formal ICRF3 position of 32 $\mu$as.  By these somewhat typical measures of variability, it would appear to be a quite stable reference frame object.  However, inspection of the time series reveals that this source has a conspicuously large coherent positional variation over $\sim$10 years, which is not adequately captured by the simple statistics. (We discuss this striking source in more detail in a forthcoming paper, Cigan et al., \textit{in prep.}) The combined (formal plus excess) error of 137 $\mu$as does better, but even if one were to attempt to determine the least variable sources by using strict filters of 0.1~mas mean jitter offset, 0.2~mas combined error, and 0.2~mas wrms -- a combination which only this source and 0059+581 satisfy, among sources with at least 10 observing sessions -- this source with clear variability would be among them.  The normalized offset metric $Q68$ and STR do successfully capture its variable nature, though, with values of 3.5 and 0.8~mas, respectively.  

The ideal reference source has the smallest possible average offset, small spread in measured data but with appropriate statistical uncertainty, and minimal drift or coherent motion over time. 
The numerous variability metrics considered in this work quantify different aspects of the variability in the astrometric measurement time series. 
The covariance-weighted mean values of the jitter offsets (Sec.~\ref{sec:jitterOffsets}) provide a robust statistical estimate of the mean position of each source, utilizing the full uncertainty and covariance information from each observing session.  
As the time series data are frequently heavy-tailed with large outliers, the formal errors of the data from the least-squares fit to the global solution of all source time series often under-represent the true variability seen in the data.  In order to capture the unmodeled error, we compute an excess error along with the jitter by incrementally adding white noise to component errors until $\chi^2_\nu$=1 is satisfied for the position measurements (Sec.~\ref{sec:ExcessError}). 
The statistical spread in the data can be assessed by a number of different methods, and a comparison of these is shown in Figure~\ref{fig:excessvariance_all_full}.  One measure we use is the common weighted root mean square of the values, a straightforward estimate of the statistical spread in, e.g., mas.  
$Q68$, the 68th percentile of the normalized position offsets $D$ (Sec.~\ref{test.sec}), is another useful measure of how noisy the observed data are. This unitless quantity is somewhat akin to signal-to-noise measurement, as position offsets are scaled by uncertainties, and would be close to a value of 1.51 for ideally distributed data and uncertainties.  Most sources have $Q68$ values slightly larger than 1.51, indicating again that most sources have offset distributions with larger tails than expected for Gaussian statistics.
Finally, we characterize the coherent motion over time with the simple STR metric introduced in Sec.~\ref{sec:STR}, where the effect of outliers is reduced and the underlying long-term drift in a source's apparent position is determined by averaging over a rolling four-month time window, with the STR being the maximum range in the smoothed time series. Even such a simple method is invaluable for capturing an aspect of the variability that the standard population statistics are not as sensitive to. 

While simple rankings of sources by single metrics are useful for determining the best and worst with regards to a specific quantity of interest, this is a limited approach since variability can manifest in several ways within these data.  
Sources can appear quite stable according to one or several metrics, and still be highly variable according to another. 
A more complete picture of which sources are most or least variable and therefore suitable for reference frame work can be obtained by combining several different criteria of variability.  
Almost all ICRF3 sources show at least moderate variability in at least one metric; strict thresholds can effectively filter out variable sources by one metric, but essentially no source falls within the top tenth percentile of all metrics explored here. 
To determine lists of the `best' sources, we employ slightly looser cuts to all considered metrics so that sources falling within the limits are those that have at least moderately low variability measures by all counts are not highly variable by any count.  
Covariance-weighted mean jitter values below 0.1~mas are a reasonable cutoff here, and since many sources have combined (formal plus excess) errors above this level, a cutoff of 0.2~mas is better.  
As many sources have $Q68$ values above the nominal $\mathcal{R}(1)$ 68-th percentile value of 1.51, a more relaxed cutoff of 3.0 selects all sources with moderate offsets but rejects extreme outliers. 
A somewhat relaxed wrms threshold of 0.3~mas was found to be reasonable in combination with the other limits without excluding the remaining good sources. 
As previously mentioned, all sources are astrometrically variable to some degree on months or years timescales, as quantified by STR (and rarely below several mas); a cutoff of 0.5~mas over the lifetime of observing histories appears to be an appropriate upper limit when used in combination with the other values.
Other rankings could of course be constructed using different cutoffs or weightings for different parameters, based on preference for which metrics to emphasize.

To assess the `worst' sources, we employ two outlooks: one is similar to that above but opposite, where moderately high but not extreme cutoffs are enforced for all metrics (these sources are highly variable in \textit{all} measures); and also one where sources that exhibit extreme variability in \textit{any} measure are captured. 
For the purpose of these rankings, lower limits for which we consider sources to be highly variable by all metrics are at least 5.0~mas for each of the mean jitter, combined error, wrms, and STR.
Of course, variability levels below these can still render a particular source unsuitable, but this combination highlights those sources which are the least stable by \textit{all} measures.
To capture the most extremely variable sources by single metrics, we use cutoffs of 10.0~mas in any one of these parameters.

Table~\ref{tab:RankedSources} lists the best and worst sources based on these rankings. The full table of values for all sources will be available in machine-readable format in the online journal. 
Again, here the best sources -- the most stable on average with combinations of good values -- are those that fall within at least moderate thresholds for all metrics, and we further restrict our consideration to only include those which have at least 30 sessions to ensure the results are more robust. 
For the full ICRF3 list of sources, those satisfying mean jitter $<$~0.10~mas, combined error $<$~0.2~mas, $Q68$ $<$~3.0, wrms $<$~0.3 mas, and STR $<$~0.5 mas are the following 10 sources: 0017+200, 0059+581, 0137+012, 0613+570, 0749+540, 1053+704, 1300+580, 1357+769, 3C371, 1806+456. 
As was discussed earlier, the sources with the longest observational histories, which are generally the ICRF3 defining sources, almost all tend to have STR values above 0.5 mas.  
Considering only the ICRF3 defining sources, the seven which satisfy all these requirements are 0017+200, 0059+581, 0613+570, 0749+540, 1053+704, 1300+580, and 1357+769.

The worst sources -- the least stable, shown to have one or more extreme variability measures -- are again broken down here by ICRF3 defining and densifying sources.
For all ICRF3 sources, 31 fall above a 5.0~mas limit in every one of the jitter, combined error, wrms, and STR parameters. 
80 extreme sources have at least one value over the much higher threshold of 10.0~mas in at least one of those parameters. 
For the defining sources to which the translation and rotation of the ICRF3 frame are aligned, eleven fall above every limit of mean jitter $>$~0.3~mas, combined error $>$~0.5~mas, wrms $>$~0.5~mas, and STR $>$~0.5~mas: 0038$-$326, 0642$-$349, 1116$-$462, 1245$-$454, 1306$-$395, 1312$-$533, 1511$-$476, 1556$-$245, 1600$-$445, 1937$-$101, and 2325$-$150.
For more extreme individual parameter outliers, 23 defining sources exceed at least one of the follow limits of mean jitter $>$~5.0~mas, combined error $>$~2.0~mas, wrms $>$~3.0~mas, or STR $>$~2.0~mas: 0013$-$005, 0038$-$326, 0308$-$611, 0316$-$444, 0539$-$057, 0742$-$562, 0855$-$716, 1027$-$186, 1034$-$374, 1116$-$462, 1143$-$332, 1406$-$267, 1435$-$218, 1451$-$400, 1600$-$445, 1642+690, 1706$-$174, NRAO530, 1754+155, 1951+355, 2002$-$375, 2220$-$351, and 2325$-$150.
More observations could help reduce the computed variability measures for some of these sources, as most have relatively short observing histories. However, the defining sources are typically observed over hundreds to thousands of sessions, and many of them still exceed these high variability thresholds.
This can be a potentially useful set of information to help inform prudent selections of future reference frame sources.

\begin{deluxetable*}{lllrrrrrrrr}
\movetabledown=0.8in
\tabletypesize{\tiny}
\tablecaption{Overall Variability Metrics by Source}
\tablehead{\colhead{B1950 Name} & \colhead{J2000 Name} & \colhead{Other Name} & \colhead{N$_\mathrm{ses}$} & \colhead{N$_\mathrm{del}$} & \colhead{$\Delta_\mathrm{mean,Tot}$} &  \colhead{$\sigma_\mathrm{ex,Tot}$} & \colhead{$\sigma_\mathrm{comb,Tot}$} & \colhead{$Q68$} & \colhead{wrms$_\mathrm{Tot}$ } & \colhead{STR$_\mathrm{Tot}$}  \\ & & & & & (mas) & (mas) & (mas) &   & (mas) & (mas) } 
\startdata 
\multicolumn{11}{c}{Defining Source List, Best Sources: }\\ 
\multicolumn{11}{c}{$\Delta_\mathrm{mean,Tot} <$ 0.1 mas, $\sigma_\mathrm{comb,Tot} <$ 0.2 mas, $Q68$ $<$3.0, wrms $<$ 0.3 mas, and STR $<$0.5 mas} \\ 
\hline
0017+200   & J0019+2021   & PKS 0017+200    &  541 &  63937 &    0.041 &    0.159 &    0.165 &     2.07 &    0.230 &    0.163 \\
0059+581   & J0102+5824   & TXS 0059+581    & 2879 & 535391 &    0.019 &    0.118 &    0.126 &     2.47 &    0.196 &    0.172 \\
0613+570   & J0617+5701   & IVS B0613+570   &  418 &  66251 &    0.002 &    0.141 &    0.148 &     2.01 &    0.261 &    0.227 \\
0749+540   & J0753+5352   & 4C +54.15       & 1179 &  86097 &    0.028 &    0.114 &    0.122 &     1.92 &    0.295 &    0.235 \\
1053+704   & J1056+7011   & S5 1053+70      &  745 &  75829 &    0.056 &    0.149 &    0.155 &     1.99 &    0.278 &    0.311 \\
1300+580   & J1302+5748   & TXS 1300+580    & 1401 & 122158 &    0.035 &    0.108 &    0.116 &     1.97 &    0.286 &    0.209 \\
1357+769   & J1357+7643   & CGRaBS J1357+7643 & 2071 & 216829 &    0.026 &    0.068 &    0.080 &     1.74 &    0.176 &    0.098 \\
\multicolumn{11}{c}{Defining Source List, Worst Sources: }\\ 
\multicolumn{11}{c}{$\Delta_\mathrm{mean,Tot} >$ 0.5 mas, $\sigma_\mathrm{comb,Tot} >$ 2.0 mas, wrms $>$ 3.0 mas, or STR $>$2.0 mas} 
\\ 
\hline
0013$-$005 & J0016$-$0015 & PKS 0013$-$00  &  114 &   3449 &    0.050 &    0.188 &    0.201 &     1.75 &    1.182 &    2.984 \\
0038$-$326 & J0040$-$3225 & PKS 0038$-$326 &   32 &    681 &    0.340 &    0.523 &    0.855 &     1.61 &    5.184 &    2.306 \\
0308$-$611 & J0309$-$6058 & PKS J0309$-$6058 & 1344 &  47810 &    0.023 &    2.189 &    2.190 &     1.99 &    0.478 &    0.403 \\
0316$-$444 & J0317$-$4414 & ESO 248$-$6    &   25 &    389 &    0.085 &    2.150 &    2.183 &     2.11 &    4.720 &    1.627 \\
0539$-$057 & J0541$-$0541 & PKS 0539$-$057 &   51 &   1844 &    0.174 &    0.439 &    0.456 &     1.94 &    3.899 &    0.796 \\
0742$-$562 & J0743$-$5619 & CGRaBS J0743$-$5619 &    8 &    176 &    1.097 &    1.073 &    1.304 &     2.68 &    0.487 &    0.536 \\
0855$-$716 & J0855$-$7149 & PKS 0855$-$716 &    8 &     82 &    0.581 &    0.681 &    0.737 &     1.48 &    0.379 &    0.420 \\
1027$-$186 & J1029$-$1852 & PKS J1029$-$1852 &  105 &   3275 &    0.121 &    0.644 &    0.650 &     1.83 &    5.351 &    1.364 \\
1034$-$374 & J1036$-$3744 & PKS J1036$-$3744 &  124 &   2922 &    0.060 &    2.121 &    2.123 &     2.26 &    1.332 &    1.431 \\
1116$-$462 & J1118$-$4634 & PKS 1116$-$462 &   44 &    541 &    0.590 &    0.954 &    0.979 &     2.35 &    1.337 &    1.663 \\
\multicolumn{11}{c}{Defining Source List, Worst Sources: }\\
\multicolumn{11}{c}{$\Delta_\mathrm{mean,Tot} >$ 0.3 mas, $\sigma_\mathrm{comb,Tot} >$ 0.5 mas, wrms $>$ 0.5 mas, and STR $>$0.5 mas} \\ 
\hline
0038$-$326 & J0040$-$3225 & PKS 0038$-$326 &   32 &    681 &    0.340 &    0.523 &    0.855 &     1.61 &    5.184 &    2.306 \\
0642$-$349 & J0644$-$3459 & PKS 0642$-$349 &   35 &   1056 &    0.464 &    1.073 &    1.159 &     2.62 &    1.854 &    1.198 \\
1116$-$462 & J1118$-$4634 & PKS 1116$-$462 &   44 &    541 &    0.590 &    0.954 &    0.979 &     2.35 &    1.337 &    1.663 \\
1245$-$457 & J1248$-$4559 & PKS 1245$-$454 &   33 &    478 &    0.361 &    0.492 &    0.547 &     1.77 &    0.939 &    0.942 \\
1306$-$395 & J1309$-$3948 & CGRaBS J1309$-$3948 &   35 &    987 &    0.454 &    0.856 &    0.906 &     1.92 &    0.618 &    0.959 \\
1312$-$533 & J1315$-$5334 & ICRF J131504.1$-$533435 &   13 &    512 &    0.318 &    1.175 &    1.213 &     3.03 &    0.845 &    0.701 \\
1511$-$476 & J1514$-$4748 & ICRF J151440.0$-$474829 &   33 &    266 &    0.397 &    0.537 &    0.616 &     2.06 &    0.622 &    0.666 \\
1556$-$245 & J1559$-$2442 & PKS 1556$-$245 &   49 &   1833 &    0.471 &    1.332 &    1.341 &     2.03 &    1.374 &    0.886 \\
1600$-$445 & J1604$-$4441 & ICRF J160431.0$-$444131 &   26 &    384 &    0.316 &    1.012 &    1.165 &     2.20 &    3.167 &    0.796 \\
1937$-$101 & J1939$-$1002 & PKS 1937$-$101 &   48 &   1526 &    0.491 &    0.737 &    0.746 &     2.51 &    1.443 &    1.165 \\
\\ 
\multicolumn{11}{c}{Full Source List, Best Sources: }\\ 
\multicolumn{11}{c}{$\Delta_\mathrm{mean,Tot} <$ 0.1 mas, $\sigma_\mathrm{comb,Tot} <$ 0.2 mas, $Q68$ $<$ 3.0, wrms $<$ 0.3 mas, and STR $<$0.5 mas} \\ 
\hline
0017+200   & J0019+2021   & PKS 0017+200   &  541 &  63937 &    0.041 &    0.159 &    0.165 &     2.07 &    0.230 &    0.163 \\
0059+581   & J0102+5824   & TXS 0059+581   & 2879 & 535391 &    0.019 &    0.118 &    0.126 &     2.47 &    0.196 &    0.172 \\
0137+012   & J0139+0131   & PKS J0139+0131 &   37 &    620 &    0.060 &    0.097 &    0.147 &     1.37 &    0.297 &    0.282 \\
0613+570   & J0617+5701   & IVS B0613+570  &  418 &  66251 &    0.002 &    0.141 &    0.148 &     2.01 &    0.261 &    0.227 \\
0749+540   & J0753+5352   & 4C +54.15      & 1179 &  86097 &    0.028 &    0.114 &    0.122 &     1.92 &    0.295 &    0.235 \\
1053+704   & J1056+7011   & S5 1053+70     &  745 &  75829 &    0.056 &    0.149 &    0.155 &     1.99 &    0.278 &    0.311 \\
1300+580   & J1302+5748   & TXS 1300+580   & 1401 & 122158 &    0.035 &    0.108 &    0.116 &     1.97 &    0.286 &    0.209 \\
1357+769   & J1357+7643   & CGRaBS J1357+7643 & 2071 & 216829 &    0.026 &    0.068 &    0.080 &     1.74 &    0.176 &    0.098 \\
1807+698   & J1806+6949   & 3C 371         & 1457 & 191284 &    0.031 &    0.063 &    0.077 &     1.95 &    0.168 &    0.142 \\
1806+456   & J1808+4542   & LB 1086        &  509 &  43701 &    0.043 &    0.151 &    0.157 &     2.01 &    0.240 &    0.230 \\
\multicolumn{11}{c}{Full Source List, Worst Sources: }\\ 
\multicolumn{11}{c}{$\Delta_\mathrm{mean,Tot} >$ 10.0 mas, $\sigma_\mathrm{comb,Tot} >$ 10.0 mas, wrms $>$ 10.0 mas, or STR $>$10.0 mas} \\ 
\hline
0008$-$421 & J0010$-$4153 & PKS 0008$-$42  &   14 &    167 &    2.570 &    9.954 &   10.012 &     5.14 &   12.930 &    6.226 \\
0022$-$423 & J0024$-$4202 & PKS 0022$-$42  &   15 &    163 &    4.823 &   16.363 &   16.369 &     5.68 &    3.590 &    9.458 \\
0030+196   & J0032+1953   & PKS 0030+19    &    8 &    108 &    3.597 &   18.607 &   18.786 &     6.65 &    7.116 &    7.110 \\
0114$-$211 & J0116$-$2052 & PKS 0114$-$21  &    9 &     78 &    6.759 &    8.931 &   10.105 &     2.41 &    2.934 &    5.227 \\
0134+329   & J0137+3309   & 3C 48          &   39 &   1681 &   56.716 &   54.225 &   54.232 &    15.77 &   13.022 &   56.894 \\
0209+168   & J0211+1707   & TXS 0209+168   &    7 &    641 &    0.128 &    0.125 &    0.275 &     1.73 &   25.337 &    0.056 \\
0304+124   & J0307+1241   & TXS 0304+125   &    5 &    234 &    7.775 &    7.773 &   10.276 &     0.88 &    0.562 &    0.257 \\
0316+162   & J0318+1628   & CTA 21         &   18 &   1431 &   10.297 &   14.031 &   14.035 &    18.86 &    4.972 &   14.326 \\
0316+413   & J0319+4130   & 3C 84          &  206 &   7546 &    1.103 &    3.319 &    3.321 &     3.62 &    6.667 &   18.967 \\
0350+177   & J0352+1754   & TXS 0350+177   &    8 &    141 &   69.224 &   74.611 &   74.679 &     9.22 &   33.901 &   10.058 \\
\multicolumn{11}{c}{Full Source List, Worst Sources: }\\ 
\multicolumn{11}{c}{$\Delta_\mathrm{mean,Tot} >$ 5.0 mas, $\sigma_\mathrm{comb,Tot} >$ 5.0 mas, wrms $>$ 5.0 mas, and STR $>$5.0 mas} \\ 
\hline
0134+329   & J0137+3309   & 3C 48          &   39 &   1681 &   56.716 &   54.225 &   54.232 &    15.77 &   13.022 &   56.894 \\
0350+177   & J0352+1754   & TXS 0350+177   &    8 &    141 &   69.224 &   74.611 &   74.679 &     9.22 &   33.901 &   10.058 \\
0709+008   & J0711+0048   & PKS J0711+0048 &    9 &     89 &   11.459 &   13.416 &   14.234 &     3.37 &    5.587 &    9.971 \\
0741$-$444 & J0743$-$4434 & ICRF J074332.7$-$443405 &   13 &    104 &    5.380 &    5.916 &    7.975 &     2.65 &    7.735 &    8.572 \\
0912$-$330 & J0914$-$3314 & PKS J0914$-$3314 &    8 &    166 &    5.006 &    9.761 &    9.824 &    12.46 &    5.182 &    7.249 \\
1015$-$314 & J1018$-$3144 & PKS J1018$-$3144 &   10 &     68 &   17.159 &   20.583 &   21.514 &     2.58 &   14.592 &   11.646 \\
1306+660   & J1308+6544   & 3C 282         &    8 &     48 &   20.190 &   26.692 &   28.494 &     1.92 &   11.813 &   12.388 \\
1320$-$446 & J1323$-$4452 & PKS 1320$-$44  &    9 &     45 &   10.761 &   71.042 &   71.094 &     9.97 &   17.874 &   33.561 \\
1323+321   & J1326+3154   & 4C +32.44      &   10 &    215 &    9.882 &    8.685 &    8.782 &     4.33 &    9.906 &   11.298 \\
1328+254   & J1330+2509   & 3C 287         &    8 &    274 &   39.816 &   72.269 &   72.275 &    50.17 &   19.710 &   47.664 \\
\\ 
\enddata
\tablecomments{
Lists of up to the first 10 ``best'' and ``worst'' sources in the full ICRF3 and defining source samples, based on numerous variability metrics. 
All sources have at least N$_\mathrm{sessions}\geq 4$, N$_\mathrm{delays}\geq 10$ per session, and a minimum span of 2 years observing history, though the best sources have a stricter requirement of 30 or more observing sessions to improve the reliability of the results. 
N$_\mathrm{ses}$ and N$\mathrm{del}$ are the number of sessions and total delays, respectively. 
$\Delta_\mathrm{mean}$ denotes the covariance-weighted mean `jitter' position offset from the ICRF3 reference coordinates. 
The designation `Tot' means that both RA and DEC components are included. 
$\sigma_\mathrm{ex}$ denotes excess error, and $\sigma_\mathrm{comb}$ denotes the combined error calculated as the quadratic sum of the excess error and the formal error.  
$Q68$ is the 68th percentile of the normalized position offsets $D$. 
The weighted root-mean-square values are denoted by wrms. 
STR denotes the smoothed time-series range.  
The full table, including R.A. and decl. components of these metrics, is published in its entirety in machine-readable format.  The ranked portion shown here illustrates its form and content.
}
\label{tab:RankedSources}
\end{deluxetable*}


\subsection{Dependence on Source Declination} \label{sec:byDEC}

\begin{figure}
\centering
\includegraphics[width=0.99\linewidth]{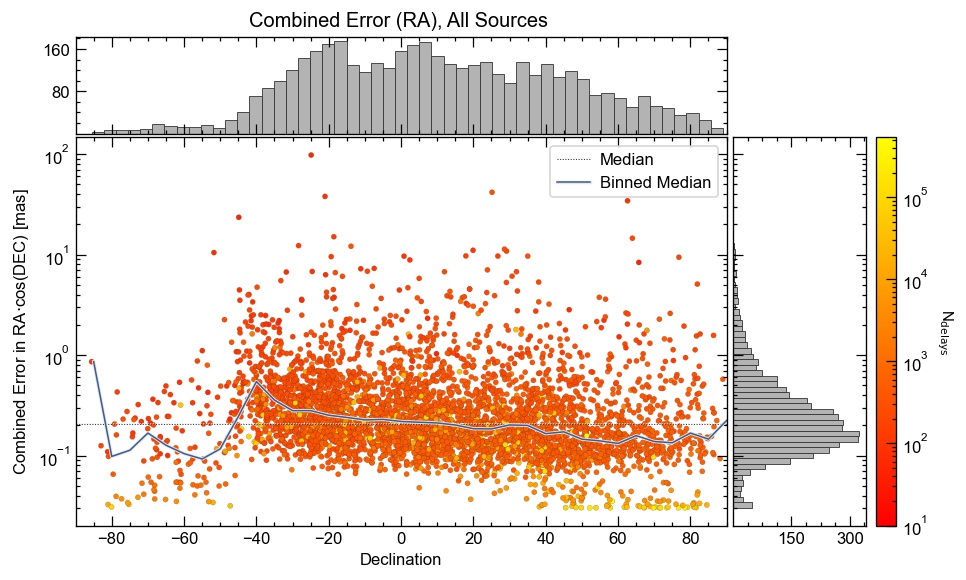}
\includegraphics[width=0.99\linewidth]{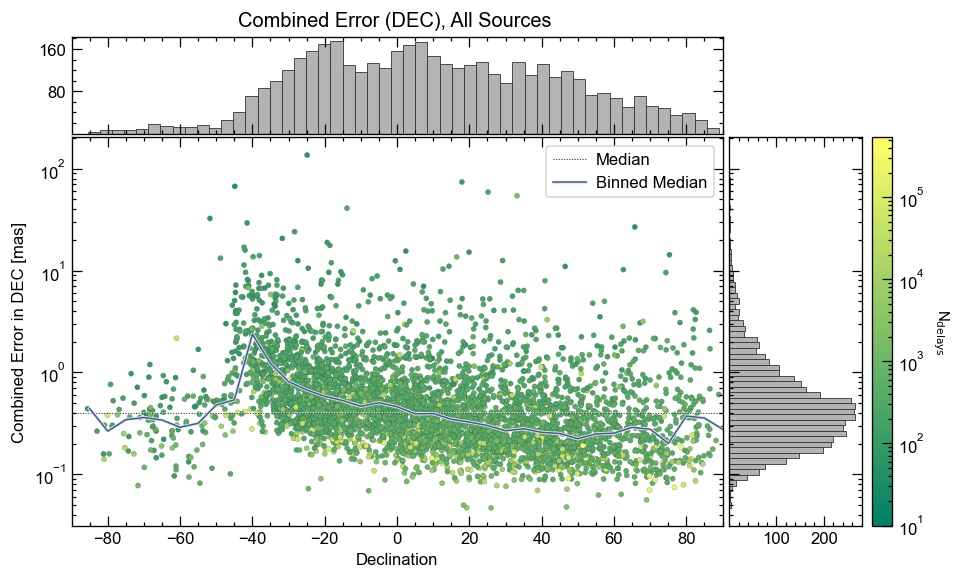}
\includegraphics[width=0.99\linewidth]{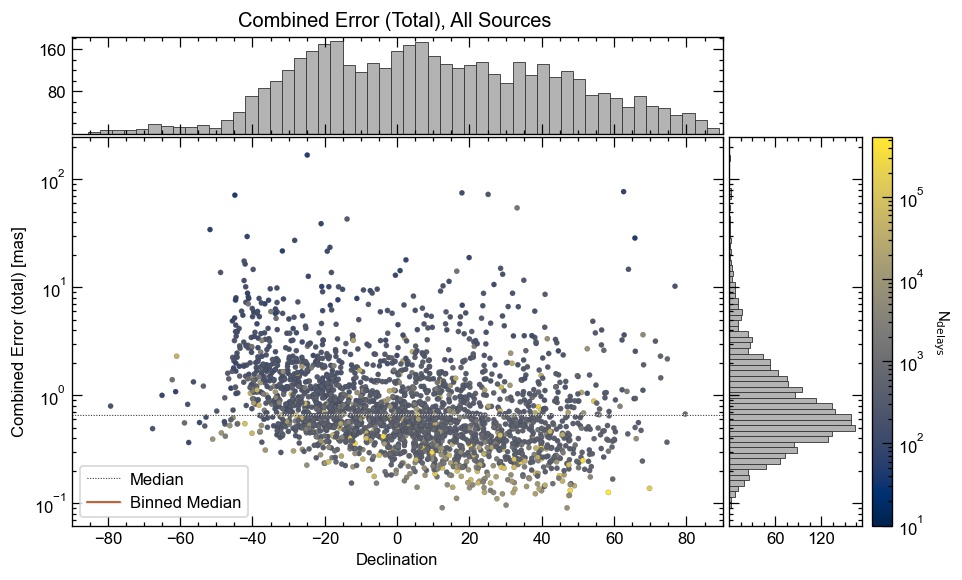}
\caption{Comparison of combined (formal plus excess) error with source declination for all ICRF3 sources with at least 4 sessions. From top to bottom, the panels show the combined error for the right ascension and declination components independently, followed by the total error (right ascension and declination components added in quadrature). Each panel shows the source data along with a rolling median in 5 degree bins, and histograms of values on the side axes. Data are color-coded by the number of observations, showing that the sources with more observations tend to have somewhat lower errors.  There is a clear increase in typical source error as declination decreases to $\sim-40\arcdeg$, owing primarily to the network of stations mostly being located at northern latitudes. The effect is most pronounced for the declination component in particular, because of baselines typically being longer in E-W than N-S directions. 
}
\label{fig:ParamsVsDec}
\end{figure}

Several VLBI-derived source position statistics exhibit features that vary with declination.  A trend of increasing formal errors as source declinations decrease to $\sim$\mbox{$-40\arcdeg$} was noted in the ICRF3 solution \citep[][their Fig. 9]{2020A&A...644A.159C}, with the R.A. errors being less pronounced than the declination errors but still present.   In our work, we also see this trend with formal errors, and moreover we see similar trends for the excess error discussed in Sec.~\ref{sec:ExcessError}.  Figure~\ref{fig:ParamsVsDec} presents the combined (formal plus excess) errors of sources as a function of their declination, by coordinate component.  The trend of increasing error as declination decreases is obvious, in particular for the declination component, with a clear cusp near $-40\arcdeg$.  This corresponds to the functional lower elevation limit of many northern stations, in particular the VLBA, which has provided a large fraction of the data comprising the global solutions. When only looking at the defining sources, the trend is even clearer. 

The visible dependence of statistical performance parameters on declination is an artifact caused by the instrument set-up and geometric configuration of the VLBI system. The majority of VLBI stations are located in the northern hemisphere. Within each session, a northern source is typically observed together with other northern sources in the phase-reference regime. This means that the baselines involved in each session are mostly east-west oriented. Since the primary measurable phase delay is the projection of the source position vector on the baseline vector, this asymmetry in the direction distribution of baseline vectors gives rise to the relative underperformance in the declination component. The geometric asymmetry, however, should be captured by the formal covariance of each single-epoch position, reflecting in the error ellipses elongated mostly in the south-north direction on the sky (Section \ref{sec:correl} and Fig. \ref{0010.fig}). There are other technical circumstances, which may not be captured by the formal uncertainties. As noted before, the dual-bandwidth X/S mode of operation is mostly employed to calibrate one of the most difficult nuisance parameters---ionospheric delay bias. This systematic error is known to be time-variable, so that the unaccounted short-term variations introduce a stochastic component of the measured delays, mostly affecting the declination component. Thus, some of the proposed astrometric variability measures may be almost free of this instrumental error overhead, while others are subject to it in full measure.

Indeed, not all variability metrics vary with declination -- $Q68$ is quite flat, as is STR.  
Other variability measures do have variation similar to the errors however, notably the wrms values of the source time series. The covariance-weighted mean jitter offsets also show a similar trend, though much less pronounced than the position errors and only for the full ICRF3 catalog; for the defining sources, which typically have high numbers of observations, the mean jitter is roughly constant with declination. 

These trends with declination are not astrophysical in origin.  The errors, jitter, and wrms were also compared to latitudes in other coordinate frames including Galactic and supergalactic coordinates.  Similar though less pronounced trends were seen in Ecliptic latitude as expected, as it is so close to Equatorial coordinates, however no significant trends were found with respect to Galactic or supergalactic latitudes.  These apparent trends with declination are primarily owing to the fact that most observed data were recorded by stations in the northern hemisphere.  Therefore more southerly sources are observed through higher columns of atmosphere, which can suffer from larger atmosphere gradients, and sources observable by these stations at their lower elevation limits tend to have increased uncertainties and variability.  Most of these trends have a cusp at $-40$ degrees in declination, which is the functional approximate lower observing limit for the northern stations.

\subsection{Looking Ahead: Implications for Reference Frames and Future Prospects} \label{sec:FuturePlans}

One immediately obvious use case for this information is for potential selection criteria in future reference frames, such as ICRF4. 
Of course, each iteration is an immense undertaking that combines the efforts of varied groups and collections of complex and evolving data sets, and many insights can only be obtained with the benefit of retrospection after the collection of more data. 
As each generation builds upon the foundations and improves upon the sometimes unavoidable limitations of the frames that preceded them, we can continue to use new analyses such as these to continue to refine and improve the reference frames upon which many global applications rely. 

Future enhancements to the software that produces the global solutions could include use of information presented here.  For example, the longer-term coherent trends highlighted by the physically-motivated time-window smoothed series could potentially be incorporated into a priori models of secular trends.  Future work into correcting for measured positions in global solutions by using models of the source structure and bulk motion could result in more precise position estimates that could translate to better reference frames, though care must be taken to keep the process from becoming circular, and ensuring enough degrees of freedom to adequately solve for all necessary parameters. 

We are currently preparing several manuscripts to further explore more aspects of these time series.  One paper will focus on several sources with conspicuous trends, playfully dubbed ``wandering quasars'' (Cigan et al., \textit{in prep.}). 
Another will investigate characteristic timescales in these data using Allan Variance and wavelet methods (Cigan et al., \textit{in prep.}), which will complement a novel robust periodogram method tested with these data \citep{2024PASP..136e4503M}.
Another project of interest using these data includes investigating the true rates of radio-optical offsets, which will build on exciting new findings from \citet{LambertSecrest2024} that blazars selected on optical photometric variability are generally more astrometrically stable objects. 

Continued observations will of course be necessary to improve the estimates of source positions and their uncertainties, in a statistical sense -- both in the legacy X/S band framework as well as reference frame observations at other frequencies and observing modes.  This new analysis provides some insight into which areas of focus could be targeted to improve specific metrics of interest to reference frame work.  Utilizing more southern hemisphere stations to observe southern sources through less atmosphere would improve the uniformity of errors across the sky, and longer N-S baselines would provide a better lever arm for determining the declination component.  The incorporation of excess (unmodeled) errors could reflect the true uncertainty in a source's series of observed positions better than the current formal errors, which are frequently underestimated.  Dedicated sessions to specifically target less-frequently observed sources and obtain time series with at least 15 sessions, for example the campaign undertaken with the VLBA at USNO, can not only help to improve the accuracy and precision of solved positions in global solutions (explored here via metrics such as the jitter and $Q68$), but also allow for more robust determinations of coherent trends over time (e.g., the STR).  A more sophisticated treatment of solved positions in global solution software, based on structure maps and information such as the 4-month window smoothed time series tracks, could potentially result in better estimates of solved source positions -- no sources are pure point sources without apparent positional variability, so incorporation of source structure models and observed secular trends where appropriate could provide improvement on standard least-squares fits of source positions. 
These of course are not simple efforts that can be quickly achieved, but if they are able to be supported they have the potential to improve our reference frames, and in turn any applications that rely upon them. Finally, we note that frequency-dependent source position, for example between emission peaks of the $S$ and $X$ bands, is an unmodeled source of error in the apparent position of the source. However, group delay-based positions, like those used for the ICRF, are less affected by these position shifts than phase delay-based positions, not being affected at all if the shift is proportional to $\nu^{-1}$ \citep{2009A&A...505L...1P}.

\section{Conclusions} \label{sec:conclusions}

Combining millions of astrometric and geodetic VLBI group delay measurements observed across 6581 sessions over the course of more than four decades of international efforts, we have produced time series of positions on the sky for over 5550 radio sources, including all 4653 sources comprising the ICRF3 X/S catalogue. This first paper in a planned series of analyses of various aspects of these time series focuses on the overall statistical aspects of the position data, with studies into characteristic timescales and other aspects of the data to to be covered in future works. Given the obvious richness and complexity of information contained within the datasets, multiple different metrics were explored that characterize distinct aspects of positional stability or variability, broadly falling into three qualitative categories.  These are:

\begin{itemize}
    \item Typical offset of single-epoch measurements (``jitter'').  For each source, we calculate this using the covariance-weighted mean, a robust estimate utilizing the full covariance information -- including RA-DEC correlations -- recorded for each single-epoch measurement. 
    
    \item Multiple estimates of the dispersion in the data, including the wrms of the position estimates, the 0.68-quantile $Q68$ of the normalized position offsets $D$, and determinations of excess variance that would need to be added to the formal errors for the observed position measurements to satisfy $\chi^2_\nu$=1.  
    
    \item A measure of the overall variation observed in coherent trends over time -- determined here by smoothing the data using a rolling four-month window in time (as opposed to fixed $N$ datapoints) and calculating the maximum position range, a value we simply dub the smoothed timeseries range (STR). 
\end{itemize}

No single metric is sufficient to adequately quantify positional stability, as all sources exhibit some form of astrometric variability with sufficient sampling in time, and combinations of different metrics should be used when gauging source quality for use in reference frame work.  We found solid statistical evidence that, whenever ICRF3 radio sources have been observed frequently enough for an extended period of time, the distribution of positional residuals based on diurnal observations exceeds the expected dispersion represented by the formal errors and their covariances. This distribution is explicitly non-Gaussian, creating a tension with the assumptions used in the computation of the weighted mean positions in the ICRF. We investigated different metrics, which quantify the degree of perturbation observed for individual sources. These metrics allow us to rank the currently observed ICRF3 sources with respect to their long-term astrometric stability, and re-prioritize the scheduling principles. Continued research into this phenomenon on both theoretical and empirical levels is warranted, given the fundamental importance of ICRF in the general construction of celestial measurements and navigation.



\begin{acknowledgments}
This work supports USNO's ongoing research into the celestial reference frame and geodesy. 
The National Radio Astronomy Observatory is a facility of the National Science Foundation operated under cooperative agreement by Associated Universities, Inc. 
The authors acknowledge use of the Very Long Baseline Array under the U.S.\ Naval Observatory’s time allocation.  
This research made use of Astropy,\footnote{\url{https://www.astropy.org}} a community-developed core Python package for Astronomy \citep{2013A&A...558A..33A, 2018AJ....156..123A, 2022ApJ...935..167A}. 
\end{acknowledgments}

%

\vspace{5mm}
\facilities{VLBA,}


\software{
    \texttt{astropy} \citep{2013A&A...558A..33A,2018AJ....156..123A,2022ApJ...935..167A},
    \texttt{ipython} \citep{ipython}, 
    \texttt{Mathematica} \citep{Mathematica}, 
    \texttt{matplotlib} \citep{matplotlib},
    \texttt{numpy}  \citep{numpy}, 
    \texttt{pandas} \citep{reback2020pandas,mckinney-proc-scipy-2010},
    \texttt{pysymlog} \citep{pysymlog} 
    }






\bibliography{sample631}{}
\bibliographystyle{aasjournal}



\end{document}